\documentclass[lettersize,journal]{IEEEtran}
\usepackage{amsmath,amsfonts,amssymb}
\usepackage{algorithmic}
\usepackage{algorithm}
\usepackage{array}
\usepackage[caption=false,font=normalsize,labelfont=sf,textfont=sf]{subfig}
\usepackage{textcomp}
\usepackage{stfloats}
\usepackage{url}
\usepackage{verbatim}
\usepackage{graphicx}
\usepackage{cite}
\usepackage[dvipsnames]{xcolor}
\begin{document}

\title{ Lens based Kinetic Inductance Detectors with Distributed Dual Polarised Absorbers for Far Infra-red Space-based Astronomy}

\author{Shahab O. Dabironezare, Giulia Conenna, Daan Roos, Dimitry Lamers, Daniela Perez Capelo, Hendrik M. Veen, David J. Thoen, Vishal Anvekar, Stephen J. C. Yates, Willem Jellema, Robert Huiting, Lorenza Ferrari, Carole Tucker, Sven L. van Berkel, Peter K. Day, Henry G. Leduc, Charles M. Bradford, Nuria Llombart, and Jochem J. A. Baselmans}

\markboth{IEEE TRANSACTIONS ON TERAHERTZ SCIENCE AND TECHNOLOGY}%
{ }

\maketitle

\begin{abstract}

Future space-based far infra-red astronomical observations require background limited detector sensitivities and scalable focal plane array solutions to realise their vast potential in observation speed. In this work, a focal plane array of lens absorber coupled Kinetic Inductance Detectors (KIDs) is proposed to fill this role. The figures of merit and design guidelines for the proposed detector concept are derived by employing a previously developed electromagnetic spectral modelling technique. Two designs operating at central frequencies of $6.98$ and $12$ THz are studied. A prototype array of the former is fabricated, and its performance is experimentally determined and validated. Specifically, the optical coupling of the detectors to incoherent distributed sources (i.e. normalised throughput) is quantified experimentally with good agreement with the estimations provided by the model. The coupling of the lens absorber prototypes to an incident plane wave, i.e. aperture efficiency, is also indirectly validated experimentally matching the expected value of $54\%$ averaged over two polarisation. The  noise equivalent power of the KIDs are also measured with limiting value of $8\times10^{-20}$ $\mathrm{W/\sqrt{Hz}}$. 
    
\end{abstract}
 

\section{Introduction}
\label{sec:I}
\IEEEPARstart{T}{he} Far-infra Red (FIR) part of the Electromagnetic (EM) spectrum contains unique information on cosmological processes which are mostly unavailable at other wavelengths, while containing a significant portion of the energy emitted within the universe \cite{Dole2006}. At this moment, no astronomical instrument targets a critical portion of this spectrum, with a wavelength range of $25-300$  $\mathrm{\mu m}$ ($1 - 12$ THz). Since Earth's atmosphere is mostly opaque for EM waves between $1$-$12$ THz, a ground-based instrument is not a feasible solution. Passively cooled space-based telescopes will reach temperatures down to $\simeq$ $50$ K, resulting in  self-emission of thermal radiation far exceeding the universe background in the FIR. Only an actively cooled ($<5$ $\mathrm{K}$) space-based telescope, in combination with extremely sensitive detectors with Noise Equivalent Power (NEP) values below $10^{-19}$ $\mathrm{W/\sqrt{Hz}}$, will be able to detect background limited radiations. Such an instrument would allow for a massive increase in spectral observing speed in the order of $\mathrm{10^5}$ with respect to existing instruments \cite{Farrah2019}. 

\begin{figure}[!t]
\centering
\includegraphics[width=3in]{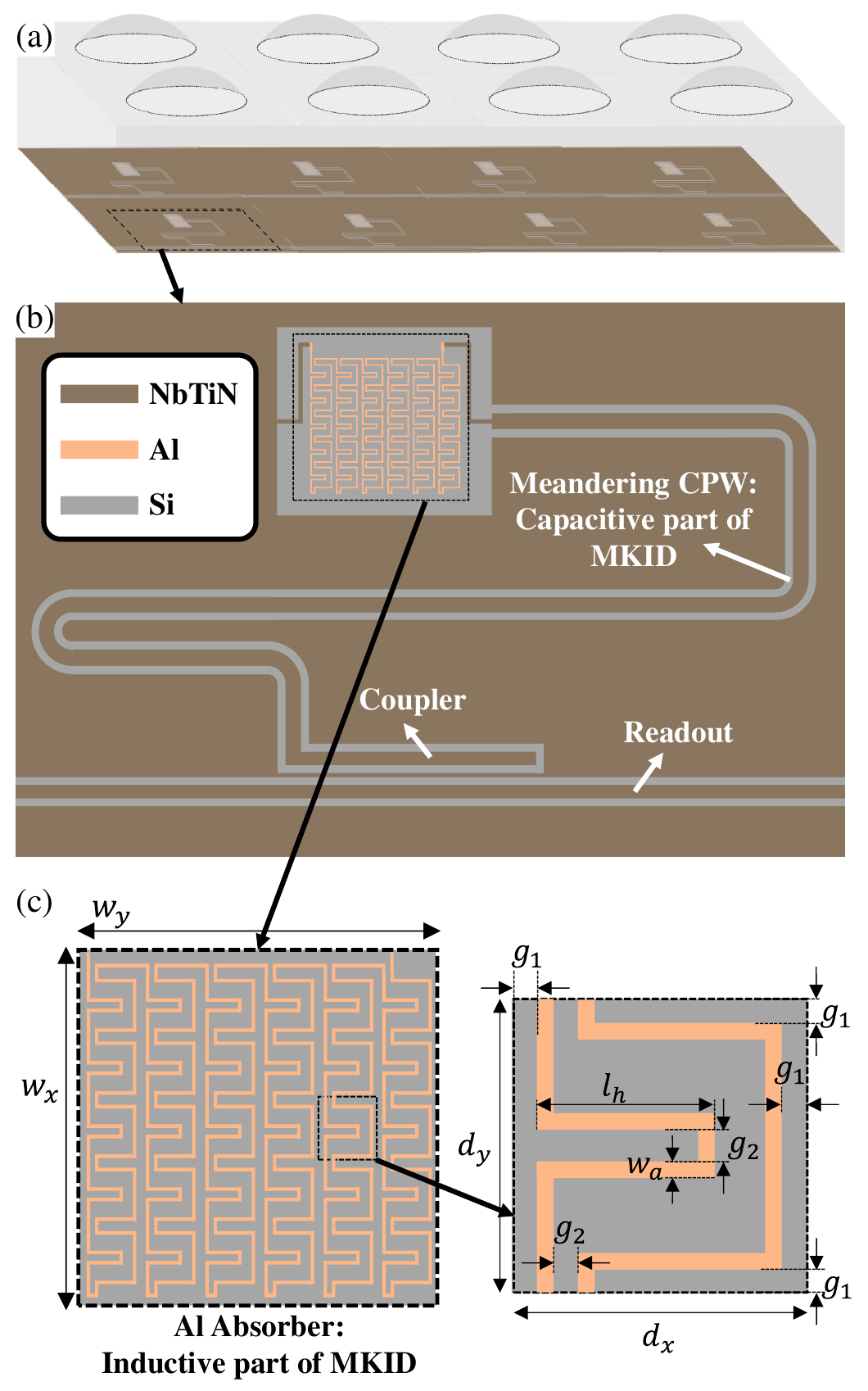}
\caption{Schematic representation of the geometry under investigation. (a) An integrated elliptical lens array with absorber coupled KIDs placed over its focal plane. (b) Zoomed in view of a detector geometry consisting of a distributed absorber connected to a quarter wavelength Coplanar Waveguide (CPW) resonator. The inductive (absorber) part of the KID is short circuited while the capacitive part (CPW) is terminated with an open circuit coupler. (c) The tightly packed periodic absorber array, its unit cell and design parameters.   }
\label{fig_geom}
\end{figure}

The PRobe far-Infrared Mission for Astrophysics (PRIMA) \cite{Glenn2024,Glenn2025} is a NASA led FIR mission aiming to close the EM observation gap between ground-based instruments such as ALMA and the space-based Mid-infra Red Instrument (MIRI) of the James Webb Space Telescope (JWST) by providing background limited radiation detection using actively cooled optics. The PRIMA observatory would contain multitude of functionalities using an imager, PRIMAger\cite{Ciesla2025}, and a grating spectrometer, FIRESS \cite{Pontoppidan2025}, for spectroscopical measurements.

Only superconductor based detectors such as Transition Edge Sensors (TES) \cite{Irwin2005}, \cite{Nagler2020}, Quantum Capacitor Detectors (QCD)  \cite{Echternach2021}, and Kinetic Inductance Detectors (KID) \cite{Day2003} are possible solutions to the extreme sensitivity requirements for PRIMA and comparable missions with actively cooled optics. Between these three archetypes of superconductor detectors KIDs have the benefits of i) allowing for large, sensitive arrays with $\sim\mathrm{10^3}$ detectors read-out using a single back-end  \cite{Baselmans2017,Bueno2018}; ii) reaching the required sensitivities, corresponding to limiting NEP values of $3\cdot10^{-20}$ $\mathrm{W/\sqrt{Hz}}$ at $1.5$ THz \cite{Baselmans2022} and $4.6\cdot10^{-20}$ $\mathrm{W/\sqrt{Hz}}$ at $12$ THz \cite{Day2024}; iii) fairly matured technology employed in stratospheric balloon instruments  \cite{Masi2019}, \cite{HD2018}; and iv) indications for tolerance against cosmic ray interactions, with a low estimated loss in integration efficiency due to cosmic rays in an orbit at the second Lagrange point of the Earth-Sun system (L2) \cite{Karatsu2019}, as well as indications that a total ionising dose for a 5 year operation in L2 has no measurable effect on device performance \cite{Karatsu2016}.

KIDs are planar superconducting resonators, capable of absorbing THz radiation, operating at a temperature $\simeq$  $\mathrm{T_c}/10$, where $\mathrm{T_c}$ is the critical temperature of the superconductor. Under these conditions all charge carriers are condensed in a Bose-Einstein condensate of Cooper pairs, which transport charge without any resistance, but with a finite inductance due to Cooper pair inertia called 'kinetic inductance'. Radiation at a frequency exceeding the gap frequency ($90$ GHz for aluminium (Al)) will break Cooper pairs into quasiparticle excitations. This process increases both the losses and the kinetic inductance \cite{Day2003}, which will change the resonant frequency and Quality (Q) factor of the resonator. A KID is therefore a microwave resonator in which an EM radiation coupling structure is embedded that allows for efficient radiation absorption. This latter is typically an antenna or absorber, whereas the rest of the resonator can be a various combination of distributed or lumped elements, creating a resonator operating typically in a frequency band of $0.5$-$8$ GHz.

In antenna based KIDs EM radiation is captured, transformed into a guided wave which propagates through a transmission line, and  finally absorbed within the resonator part of the KID. In contrast, in an absorber based KID, EM radiation is directly absorbed at the EM sensitive part of the detector which is also part of the resonator. Antenna based KIDs, have a better coupling control to incoming EM radiation due to their spatial filtering capability \cite{Llombart2018}. On the other hand, specifically at high THz ($> 5$ THz) frequencies, an absorber based KID, removes the complexities of coupling to a THz transmission line. Moreover, the fabrication and assembly tolerances are significantly relaxed in such a detector with respect to the antenna case, since former is an incoherent detector and not sensitive to the phase of the incoming radiation.   

Distributed bare absorbers have been employed as lumped element KIDs in instruments such as \cite{HD2018}, \cite{Calvo2016} and \cite{Barry2018} where the inductive part of the KID's resonator doubles as the radiation absorbing structure. By introducing a focusing dielectric lens component on top of the absorbers as shown in Fig. \ref{fig_geom}(a), the size of the absorber can be reduced, limited to the focal spot of the lens, leaving room to freely design the resonator's capacitive section without reducing the array fill factor. Moreover, the  presence of a dielectric lens on top of the absorber acts as an additional spatial filter which reduces the sensitivity of the detector to stray light absorption with respect to a bare absorber design. 

 In this work we present the design and experimental validation of lens-absorbers based KIDs. We employ a spectral model to co-design the absorbers and lenses, resulting in a very efficient and fast design process. Such a design is simply not achievable resorting only to full wave simulators. This method is built upon the work in \cite{Llombart2015}; derived in \cite{Llombart2018} for bare absorbers below a parabolic reflector, in which we drew comparisons to antenna based systems and showed detailed validations against full wave simulations; and experimentally validated for an bolometer in \cite{Dabironezare2018}. The technique can be readily extended to entirety of an imaging system via \cite{Dabironezare2021} to optimise the observation speed of an instrument as investigated in \cite{sven2025}. Our design employs a unit cell with two parallel meandering narrow Al lines which provides a wide operational bandwidth, high coupling efficiency, and dual polarised response. We focus on designs at two frequency bands, relevant to the PRIMAger instrument, centred at $6.98$ and $12$ THz. Furthermore, the end-end performance of the $6.98$ THz design is evaluated by fabricating a 25-KID array and measuring their NEPs and optical couplings to an incoherent distributed source (also referred to as the normalised throughput in the literature \cite{Griffin2002}, \cite{Llombart2018}). 

The paper is structured as follows. Section II contains guidelines to co-design distributed absorber structures below integrated lens components with a detailed performance study over the relevant design parameters. In Sec. III, the designed lens absorber coupled KID geometries are presented. Section IV is focused on the experimental characterisation of the fabricated devices and comparison to the ones estimated by the model. Concluding remarks are provided in Section V. The EM figures of merit of a lens absorber are reviewed in Appendix \ref{App_fig_mert}. The additional data regarding the EM performance of the designed lens absorbers are provided in Appendix \ref{App_EM}. The analytical expressions for designing the coplanar waveguide (CPW) section of the KIDs are provided in Appendix \ref{App_CPW}. The experimental setup is provided in Appendix \ref{App_setup}. In Appendix \ref{App_opt_coup}, the steps taken to model and experimentally quantify the optical coupling for a multi-mode detector is formulated. The additional experimentally obtained data are given in Appendix \ref{App_Result}. In the accompanying supplementary document, a tolerance study, relevant for development of detectors at FIR wavelengths, is provided on lens absorber geometries.

\section{Lens Absorber Design Guidelines}
\label{sec:II}

Based on a previously developed spectral modelling technique, \cite{Llombart2015} and \cite{Llombart2018}, this section is focused on the co-design guidelines for lenses and distributed absorbers. In this paper for brevity, the latter is referred to as absorber. The guidelines provided in this section are employed to design and evaluate the performance of the lens absorber prototype in Sec. \ref{sec:III}. 
   
\begin{figure}[!t]
\centering
\includegraphics[width=3in]{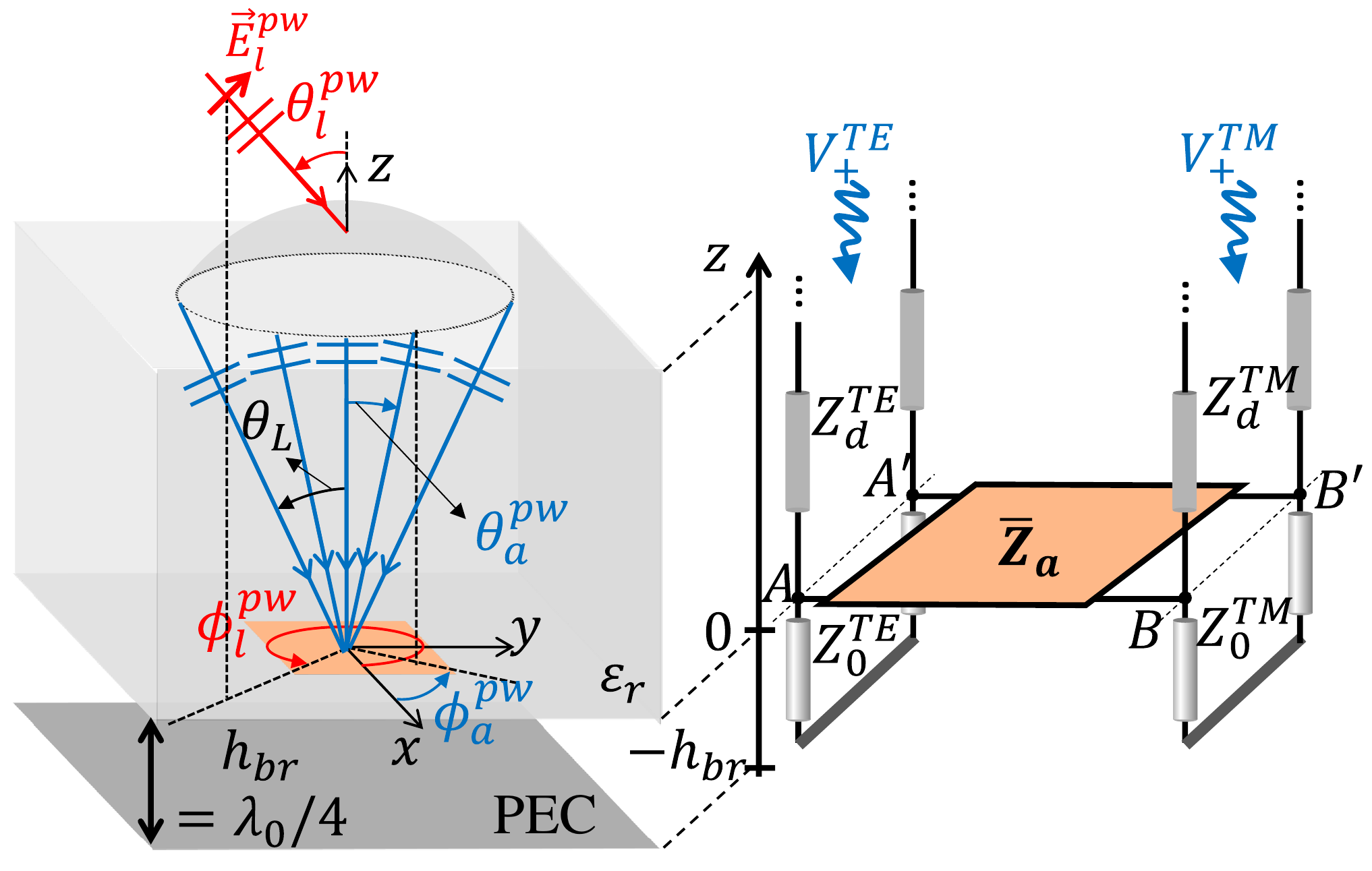}
\caption{An example lens absorber geometry with quarter wavelength backing reflector (left-hand side) and its equivalent Floquet wave model (right-hand side). The lens is illuminated by an incident plane wave from $\theta_{l}^{pw}$, $\phi_{l}^{pw}$ direction, its subtended angle is limited to $\theta_L$, and its focal field is model as a PWS. The absorber is modelled as a two-port impedance network containing its spectral response to plane waves arriving from each $\theta_{a}^{\mathrm{pw}}$, $\phi_{a}^{\mathrm{pw}}$ combination of angles.   }
\label{fig_Zmat} 
\end{figure}

Detector arrays for a FIR space observatory will have multiple possible functionalities such as spectroscopy, polarimetry, and imaging \cite{Glenn2025}. The lens Focal Plane  Arrays (FPAs) will be typically placed below reflector chains with a large ($>10$) equivalent focal to diameter (f-number) ratio.  Large f-number requirements and tight sampling for the lens FPA permits us to consider a study in which an incident plane wave directly illuminates the lens FPA, see Fig. \ref{fig_Zmat}, and to focus on demonstrating the performance of lens absorber components in isolation. As a result, the considered design parameters here are not the lens diameter (sampling rate below the reflector system), nor the design of the reflector chain and its equivalent f-number. Such parameters can be designed for specific observation scenarios based on guidelines in \cite{sven2025}. Instead, here we focused on the lens f-number, absorber size, absorber unit cell design, and KID design.

\subsection{EM Response of a Lens Absorber to Incident Plane Waves}
\label{sec:IIA}

Our absorber design consist of a finite array of lossy unit cells. The absorption response of a tightly periodic array of unit cells to an incident plane wave can be estimated via the fundamental Floquet wave modes while neglecting finiteness effects. In this work, this response is referred to as the \textit{spectral response} of an absorber. This response can be derived analytically for simple geometries such as resistive sheets and strip absorber arrays, as in \cite{Llombart2018} and \cite{Bea2014}, respectively. For arrays with a complex unit cell geometry, one can obtain this response numerically using the periodic boundary conditions in a full wave simulator, e.g CST MS \cite{ref_CST}, as in \cite{Dabironezare2018}. The Floquet wave model is shown in Fig. \ref{fig_Zmat} for an incident plane wave illuminating the absorber from colatitude and azimuth angles, $\theta_{a}^{pw}$ and $\phi_{a}^{pw}$, respectively. The model contains an equivalent transmission line with two ports and generators, Transverse Electric (TE) and Transverse Magnetic (TM) fundamental Floquet modes, representing the stratification and incident plane waves. The absorber's response to is included as a two-port impedance network, $\bar{Z}_a$. 


The response of a focusing Quasi-Optical (QO) component to an incident plane wave, i.e. the focal field generated from this plane wave, can be represented as a summation of plane waves, referred to as direct Plane Wave zpectrum (PWS) \cite{Llombart2015}. The term \textit{direct} indicates that the field distribution is obtained without the presence of the absorber. This PWS is limited to the focusing component's maximum subtended angle, $\theta_L$, see Fig. \ref{fig_Zmat}. The field distribution with inclusion of the absorber's spectral response is referred to as the \textit{total} PWS. This PWS is constructed by solving the described Floquet wave circuit per spectral direction. By coherently summing the total PWS, the total spatial fields at the area containing the absorber are obtained. The Poynting vector associated to these fields is then integrated over the absorber's physical domain to calculate the power absorbed \cite{Llombart2018}:

\begin{equation}
\label{Eq_P_abs}
	\begin{split}
 & P_{\mathrm{abs}}(f,\theta_{l}^{\mathrm{pw}},\phi_{l}^{\mathrm{pw}}) = \\ & \dfrac{1}{2}\Re{ \Big\{\iint_{-w/2}^{w/2}{[\vec{e}_t(\vec{\rho},\theta_{l}^{\mathrm{pw}},\phi_{l}^{\mathrm{pw}})\times\vec{h}_t^*(\vec{\rho},\theta_{l}^{\mathrm{pw}},\phi_{l}^{\mathrm{pw}})]\cdot\hat{z} \;\mathrm{d}\vec{\rho}}\Big\}} 
 \end{split}
\end{equation}

\noindent 
where $f$ is the frequency of operation, $\theta_{l}^{\mathrm{pw}}$ and $\phi_{l}^{\mathrm{pw}}$ are the colatitude and azimuth angles, respectively, of the incident plane wave \textit{illuminating the lens}, $w$ is the side length of an square domain absorber, $\vec{e}_t$ and $\vec{h}_t$ are the total electric and magnetic spatial fields, respectively, evaluated over the focal plane of the lens at 
 $\vec{\rho}=x\hat{x} + y\hat{y}$ locations. Eq. (\ref{Eq_P_abs}) quantifies the amount of power the absorber receives from point sources located at $\theta_{l}^{\mathrm{pw}}$ and $\phi_{l}^{\mathrm{pw}}$ angular location outside the lens. Relevant figures of merits, derived in \cite{Llombart2018} for a bare absorber below an equivalent parabolic reflector, are briefly reviewed in Appendix \ref{App_fig_mert} for lens absorbers.

\subsection{Co-design of the Lens-Absorber}
\label{sec:IIB}

\begin{figure}[!t]
\centering
\includegraphics[width=3in]{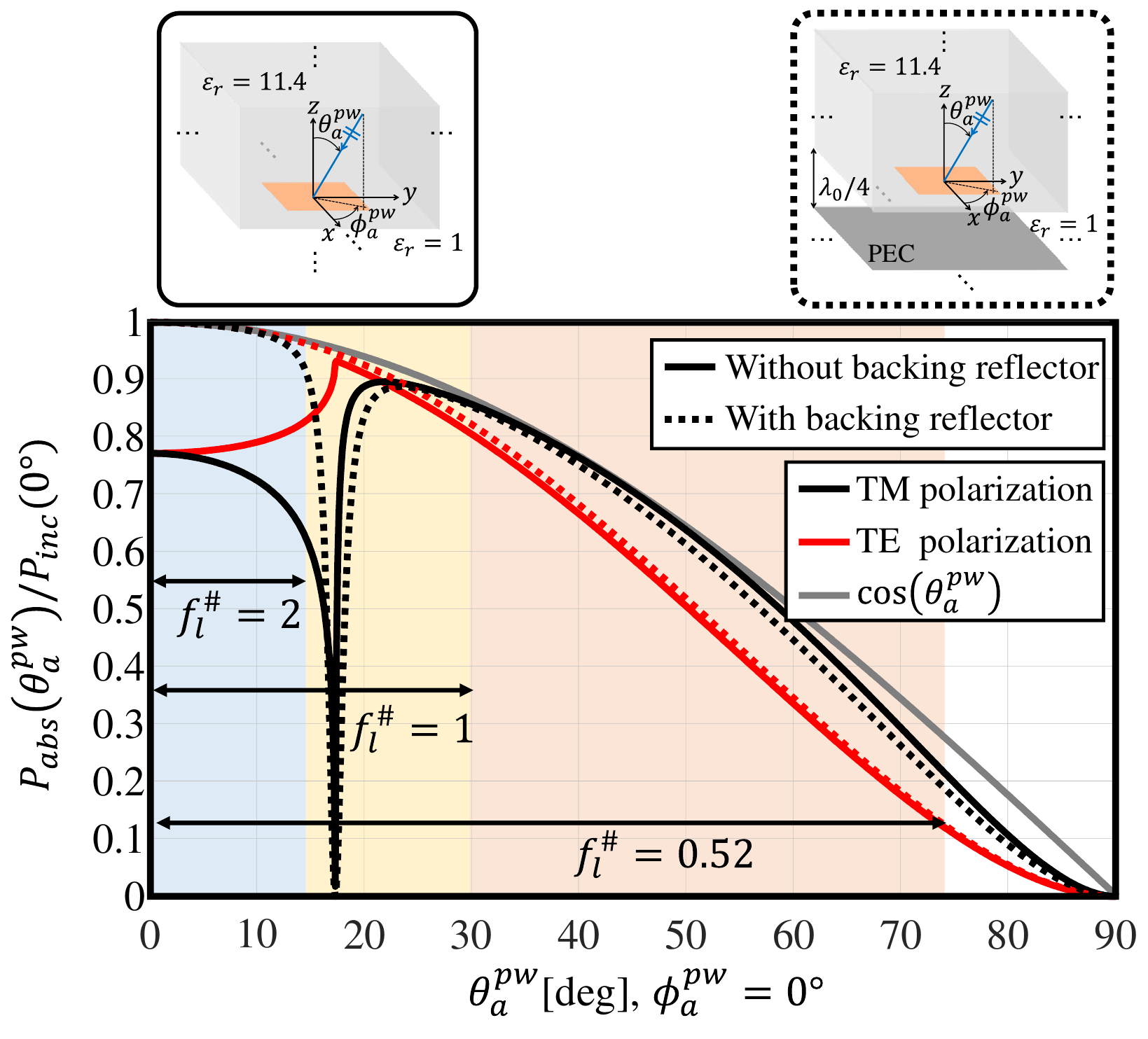}
\caption{Spectral response (absorption rate) of ideal absorbers with or without quarter wavelength backing reflector are shown as a function incident plane wave angles. These plane waves are propagating within a semi-infinite silicon slab and the absorber is located at air-silicon interface. The angular regions corresponding to the three considered lens f-number cases are also marked.  }
\label{fig_PW_id_abs}
\end{figure}

Let us consider two idealised absorber geometries coupled to silicon integrated lenses to investigate the trade-offs present in such geometries. i) An absorber with equivalent sheet resistance of $R_s= \zeta_d || \zeta_0 $ $\Omega/ \Box$ placed at the air-silicon interface, where $\zeta_d$ is the characteristic impedance of the high-resistivity silicon at cryogenic temperatures (with permittivity of $\varepsilon_r=11.44$) and $\zeta_0$ is the impedance of free space; ii) an absorbing structure with  $R_s= \zeta_d $ $ \Omega/ \Box$ placed at the air-silicon interface above a quarter wavelength backing reflector. The spectral response of these two ideal absorbers is shown in Fig. \ref{fig_PW_id_abs}, similar to the one in \cite{Llombart2015}, with the insets indicating their stratification. The response of both is limited to the cosine law of Lambert, while the cases without or with the backing reflector have maximum broadside absorption rates of $77\%$ and $100\%$, respectively. In both stratification cases, a null exists in the TM polarised response at the critical angle valued corresponding to the air-silicon interface. 

\begin{figure}[!t]
\centering
\includegraphics[width=3in]{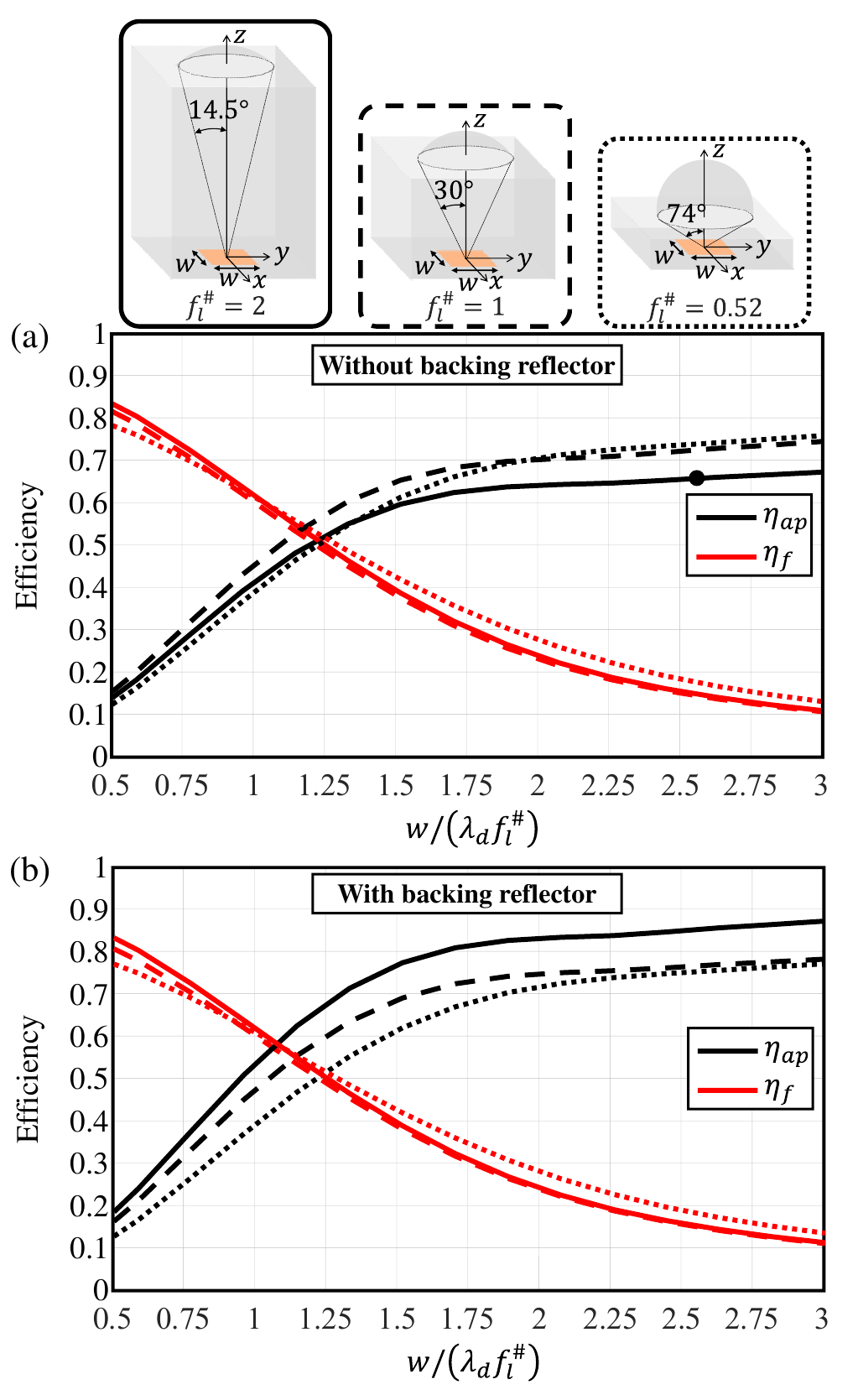}
\caption{Aperture and focusing efficiencies as a function of absorber sampling rate for three lens f-number cases: ideal absorber (a) without, and (b) with quarter wavelength backing reflector. The example incident plane wave illuminating the lenses is y-polarised. Solid, dashed and dotted lines correspond to lens f-numbers of $2$, $1$, and $0.52$, respectively. The top insets indicate these lens geometries. The black coloured circle in (a) represents the theoretical limit of the aperture efficiency for the fabricated prototype.   }
\label{fig_lens_absorber_ideal_eta} 
\end{figure}

In Fig. \ref{fig_PW_id_abs}, the angular regions subtended by three example lens elements are also marked. Here silicon lenses with three f-numbers are considered: i) shallow truncated elliptical lens with $f_{l}^{\#}=2$ and $\theta_L=14.5^{\circ} $; ii) a moderately  truncated lens with $f_{l}^{\#}=1$  and $\theta_L=30^{\circ}$; iii) an untruncated ellipsoid with $f_{l}^{\#}=0.52$  and $\theta_L=74^{\circ} $. 
The lens absorber's performance in terms of aperture and focusing efficiencies are plotted in Fig. \ref{fig_lens_absorber_ideal_eta} by varying the absorber side length for the three considered elliptical silicon lenses. Standard quarter wavelength anti-reflection coatings are considered on the lenses.
 
A constant and equal to one spectral response, instead of the ones shown in Fig. \ref{fig_PW_id_abs}, for an absorber within its lens subtended angle leads to a total PWS equal to the direct PWS. A focal field constructed from coherently summing the direct PWS have the narrowest possible main beam which approaches the one of the Airy pattern for lenses with large f-numbers. Such a narrow focal field in turn is equivalent to absorbing more power from the intended source direction with a fixed absorber size. However, a unitary spectral response is not feasible due to the cosine law but approaching it is the design goal of absorber unit cells. The enhancement of the aperture efficiency for an ideal absorber with backing reflector below a lens with $f_l^{\#}=2$, solid black coloured curve in Fig. \ref{fig_lens_absorber_ideal_eta}(b), showcases this goal since the spectral response of this absorber is relatively unitary within the angular region of its lens, see Fig. \ref{fig_PW_id_abs}. In contrast, for cases without backing reflector, lenses with larger subtended angular regions achieve higher aperture efficiencies due to the presence of a maximum in the absorber's spectral response outside the angular domain of the $f_l^{\#}=2$ lens.  

In terms of the angular selectivity of the lens absorber, modifying the lens f-number and absorber's spectral response leads to minute variations while the absorber's size dominates the behaviour. This is shown in Fig. \ref{fig_lens_absorber_ideal_eta} by comparing the focusing efficiencies for ideal absorbers below lenses with different f-numbers and spectral responses. In Appendix \ref{App_fig_mert} Fig. \ref{fig_lens_absorber_ideal_PRx}, the same behaviour is shown in terms of widening of the normalised reception power pattern when larger (in terms of $\lambda_d f_l^{\#}$, where $\lambda_d$ is the wavelength in silicon) absorbers are placed below a lens.

To summarise, the trade-offs between focusing and aperture efficiencies dictates a specific design guideline: the absorber's spectral response and the lens angular domain should be co-designed to achieve the highest aperture efficiency using the smallest possible absorber physical size. Such an approach leads to maximising the power received from the intended source direction (high $\eta_{ap}$), and reducing the detector's sensitivity to stray light and sky background radiations (high $\eta_{f}$), i.e. increasing the signal power while receiving the smallest amount of background noise.  
 
It is worth emphasising that an approach similar to the one discussed in this section is feasible for analysing and co-designing a complete lens absorber FPA coupled to a reflector QO system. This can be achieved by obtaining the direct PWS of the cascaded focusing system using the formulations derived in \cite{Dabironezare2021} by combining a numerical Geometrical Optics technique with the Coherent Fourier Optics approach. 

\section{Detector Design}
\label{sec:III}

 In this section we present the design of two lens-absorber geometries operating at central frequencies of $6.98$ and $12$ THz. We provide their design parameters and estimate their performance in terms of aperture efficiency and reception power pattern. In subsections \ref{sec:IIIA} we present the absorber's unit cell design and spectral response, and in subsection \ref{sec:IIIB} the lens-absorber design. In subsections \ref{sec:IIIC}  we discuss the integration of the absorber into a microwave resonator to complete the KID design. 

\subsection{Absorber Unit Cell Design}
\label{sec:IIIA}

\begin{figure}[!t]
\centering
\includegraphics[width=3in]{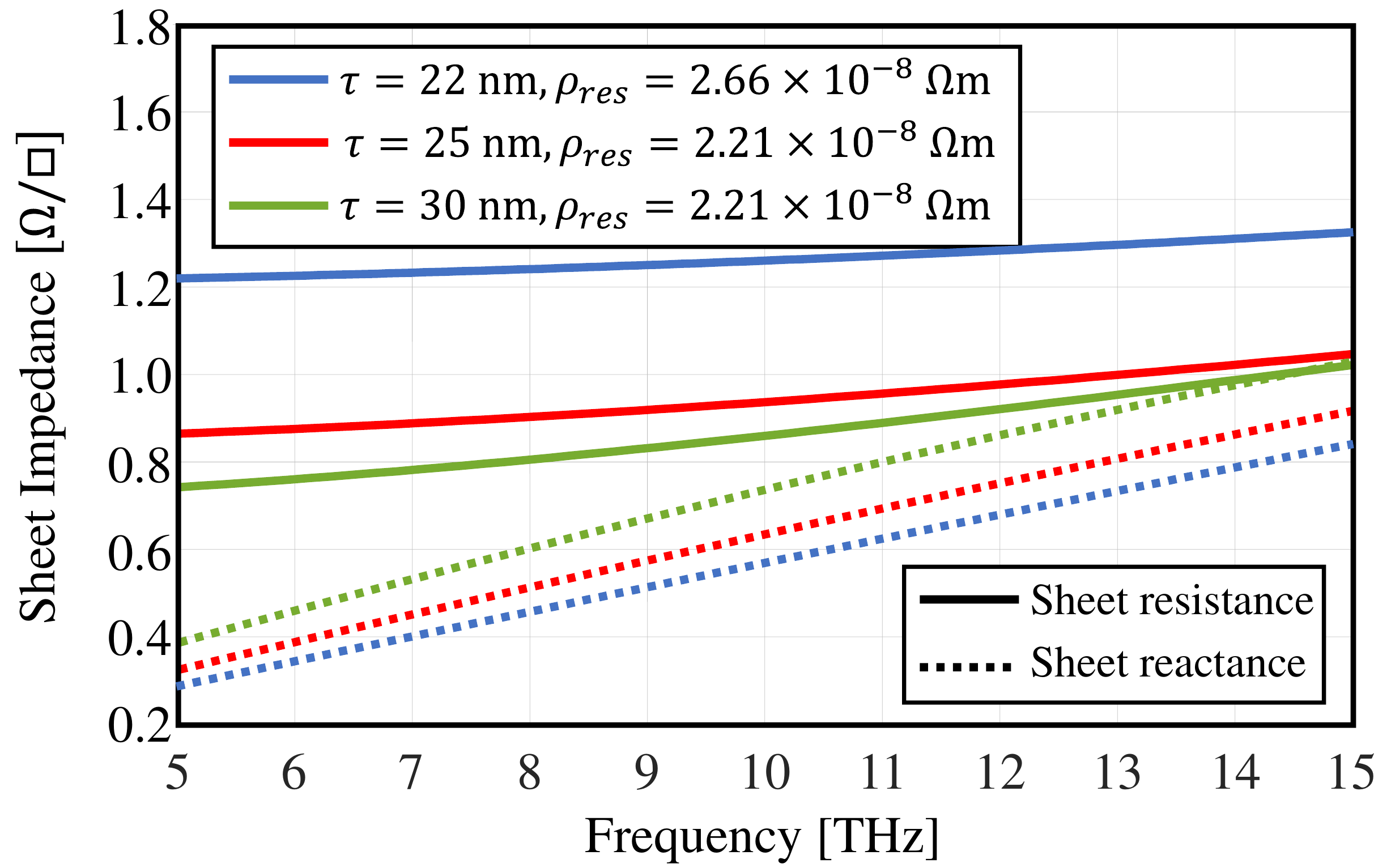}
\caption{Sheet impedance of example Al films deposited on silicon substrate with $\tau$ thickness and DC  resistivity of $\rho_{res}$.  }
\label{fig_sheet_imp} 
\end{figure}

We first study the properties of Al films. In Fig. \ref{fig_sheet_imp} we show the complex surface impedance of three Al films with thickness of $\tau$ and resistivity of $\rho_{res}$, using expressions given in \cite{Matick1995} and the measured DC resistivity at $1.5$ K given in the legend. As can be seen, the sheet reactance of Al cannot be neglected with respect to its resistance at FIR wavelengths.  As a result, a standard strip absorber design, as in \cite{Bea2014}, will have a significant reactance and will be practically impossible to match to the (real) impedance of the FIR radiation on the Si-air interface. Even matching the real part of the impedance of such an absorber is not feasible in practice: it would require extremely narrow ($<100$ $\mathrm{nm}$) and thin ($<10$ $\mathrm{nm}$) strips, which are unattainable in fabrication and would significantly reduce the maximum microwave readout power handling of the KIDs. In this work, we propose a unit cell consisting of two meandering Al strips, explored in \cite{Molero2017} to develop dual polarised absorbers, which is employed here to achieve relatively wide band impedance matching at THz frequencies using Al. The proposed design is shown in Fig. \ref{fig_geom}(c). The advantage of this design with two meandering strips over a single one \cite{Day2024} is a wider bandwidth, which additionally results in an increased tolerance to variations in the width of fabricated lines as well as variations in the resistivity of Al films.

\begin{figure}[!t]
\centering
\includegraphics[width=3in]{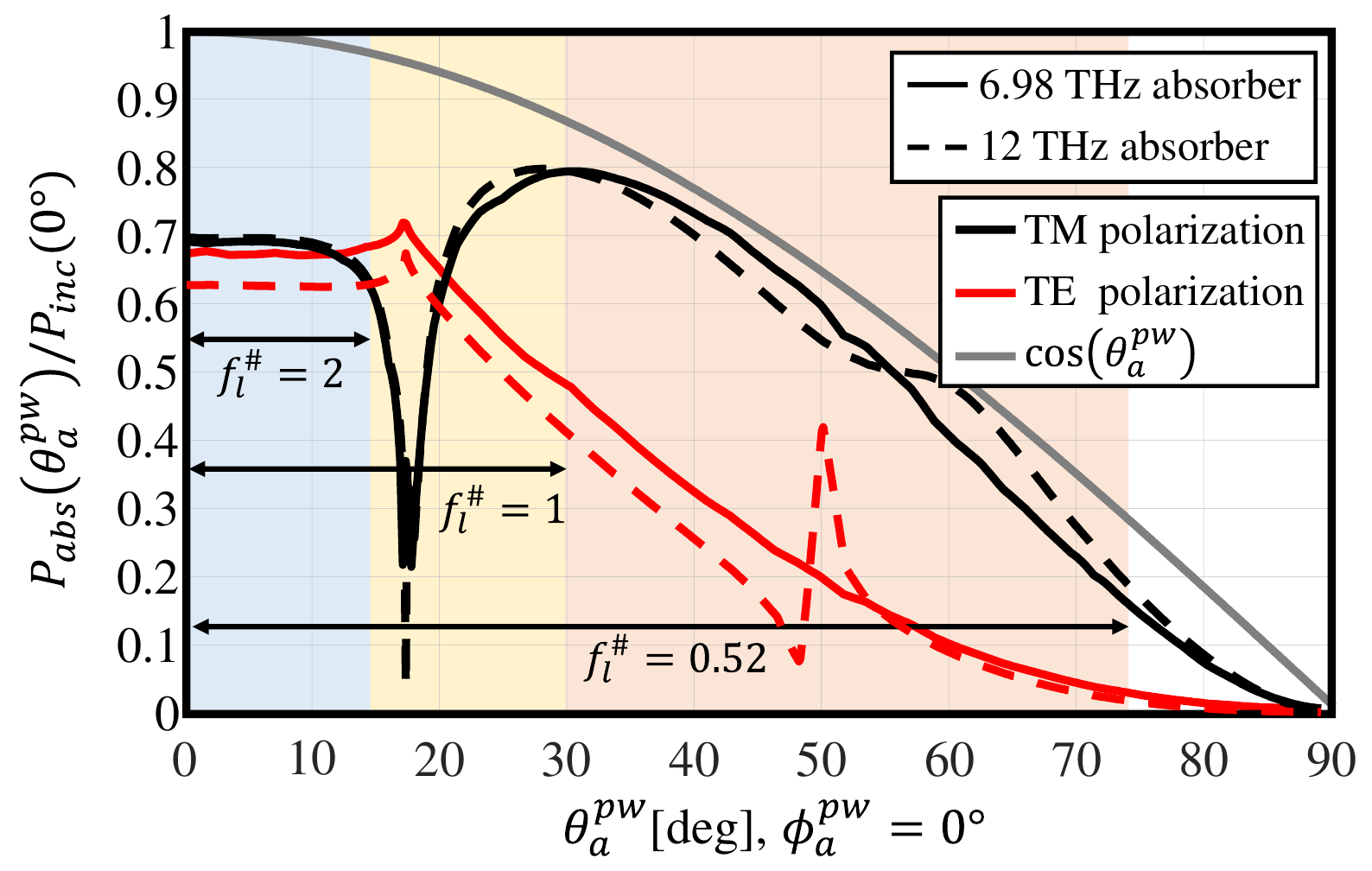}
\caption{ The spectral response of the proposed absorber unit cells to incident plane waves at  $\phi_a^{\mathrm{pw}}=0 ^{\circ} $ cut. The indicated angular regions correspond to the three considered lens f-number cases.  }
\label{fig_PW_Shabs_vs_ang} 
\end{figure}

By optimising the geometrical parameters of the proposed unit cells, one can quasi-independently tune the impedance of the structure seen by  TE and TM polarised plane waves. Such a structure has a resonant behaviour due to the combination of the capacitive and inductive responses of a dual meandering strip \cite{Molero2017} with the Al sheet impedance. As a result, it provides a large degree of freedom to tune the frequency response of the absorber. We have concluded from the optimisation process that narrower and thinner Al strips lead to wider operation bandwidths. However, both of these requirements lead to a more challenging fabrication process. As a result, the Al thicknesses chosen for $6.98$ and $12$ THz absorber designs are $22$ $\mathrm{nm}$  with a DC resistivity of about $2.66 \times 10^{-8} $ $\Omega$m.

 The geometrical parameters of the unit cells operating at the two frequency bands are reported in Table \ref{tab_des_para}. The absorbers discussed here are designed without the inclusion of a quarter wavelength backing reflector to reduce the fabrication complexity of the initial prototypes. The spectral response of these absorbers at the centre of their operation band is shown in Fig. \ref{fig_PW_Shabs_vs_ang}. 
 A tight periodicity was enforced in the designs to limit the absorber's spectral response to only fundamental TE and TM Floquet waves. As a result, scan blindness does not occur over the operation bands and within the angular domain subtended by the truncated elliptical lenses. By employing lens elements with limited angular domains, extreme tight periodicity requirements on the absorber unit cell could be relaxed. 
  For the described designs, the presence of a higher order TE Floquet wave mode is noticeable in Fig. \ref{fig_PW_Shabs_vs_ang} for the $12$ THz absorber at angles above $50^{\circ}$. However, the influence of this mode on the performance of the absorber below an untrancated lens is negligible due to the damping of the response by the cosine law at large skewed angles.
  
\begin{table}[]
\caption{Design parameters of the two absorber geometries given in $\mathrm{\mu m}$}
\label{tab_des_para}
\resizebox{3.5in}{!}{%
\begin{tabular}{|c|c|c|c|c|c|c|c|c|}
\hline
       Absorber type 	& $w_x$	& $w_y$ 	& $d_x$	& $d_y$	& $w_a$	& $l_h$	& $g_1$	& $g_2$	\\ \hline
       
		6.98 THz		& 64		& 61.6	& 6.4	& 7.7	& 0.3	& 3.8	& 0.3	& 0.5	\\ \hline

		12 THz			& 38.7	& 38.4	& 4.3	& 4.8	& 0.25	& 2.3	& 0.25	& 0.5	\\ \hline

\end{tabular}%
}
\end{table}

\subsection{Lens Absorber Design}
\label{sec:IIIB}
 
The lens f-number and absorber array's side length were co-designed to satisfy the following criteria: experimentally demonstrating the aperture efficiency estimated by the model while reducing the risks associated with realisation inaccuracies for the initial prototype. Consequently, a tolerance study was performed to estimate how sensitive is the aperture efficiency to four relevant issues at FIR wavelengths: i) curvature inaccuracies in fabrication of the elliptical lens surface; ii) surface roughness of the lens; iii) lateral and vertical misalignments between the centre of the absorber array and the lens lower focus; and iv) the presence of air gaps in the silicon wafer stacks. This study is provided in the supplementary document accompanying the paper.

Based on the conclusions of the tolerance study and the estimated performance of lens absorber designs, together with the requirements for PRIMAger, lens elements with f-number $\simeq 2$ and a diameter of $D_l = 700$ $\mathrm{\mu}m$ were chosen. Moreover, absorbers with side lengths of $w\simeq 2.55 \lambda_d f_l^{\#}$ are fabricated due to their tolerance to alignment errors. The  absorber side lengths, in $\mathrm{\mu}$m, are also provided in Table \ref{tab_des_para}.

\begin{figure}[!t]
\centering
\includegraphics[width=3in]{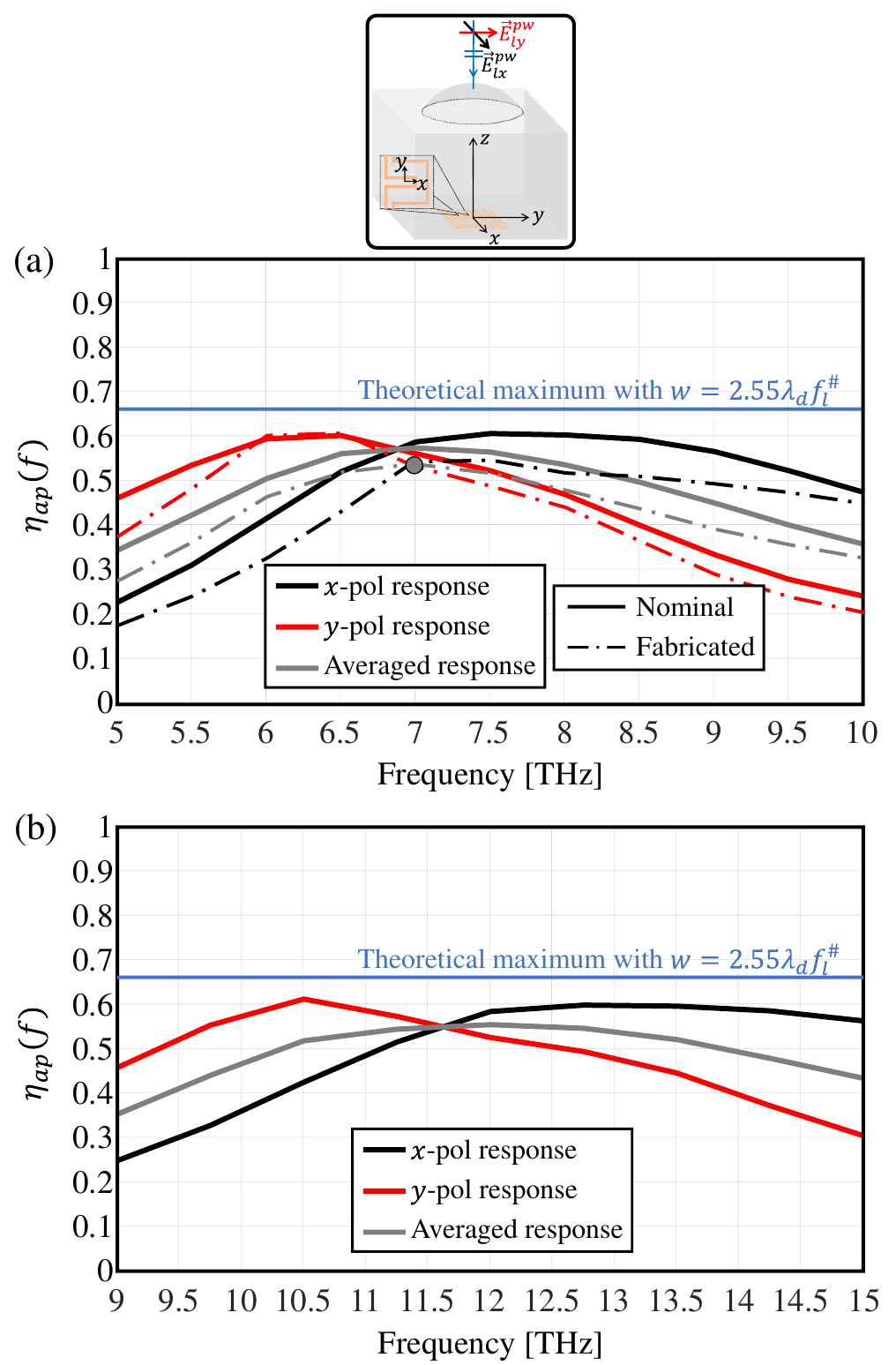}
\caption{The aperture efficiency of the designed (a) $6.98$ and (b) $12$ THz lens absorbers against their frequency of operation when the lens is illuminated by a $x$- or $y$-polarised plane wave from broadside direction. The averaged response to a combination of the two linear polarisations is also provided. The solid and dotted lines correspond to the designed and fabricated devices, respectively. The blue solid lines indicate the theoretical maximum value based on Fig. \ref{fig_lens_absorber_ideal_eta}(a). }
\label{fig_ap_eff_LensShabs} 
\end{figure}

Lens absorber performance in terms of aperture efficiency over their operation frequency bands is reported in Fig. \ref{fig_ap_eff_LensShabs} . The aperture efficiency of the two designs averaged over the two linear polarisations exhibits a $1$ dB bandwidth of $3.2$ and $5.1$ THz ($45.8 \%$ and $42.5 \%$ relative bandwidth) and a maximum aperture efficiency $10\%$ below the theoretical maximum for $6.98$ and $12$ THz designs, respectively. 

\subsection{KID Design}
\label{sec:IIIC}

\begin{figure}[!t]
\centering
\includegraphics[width=3.2in]{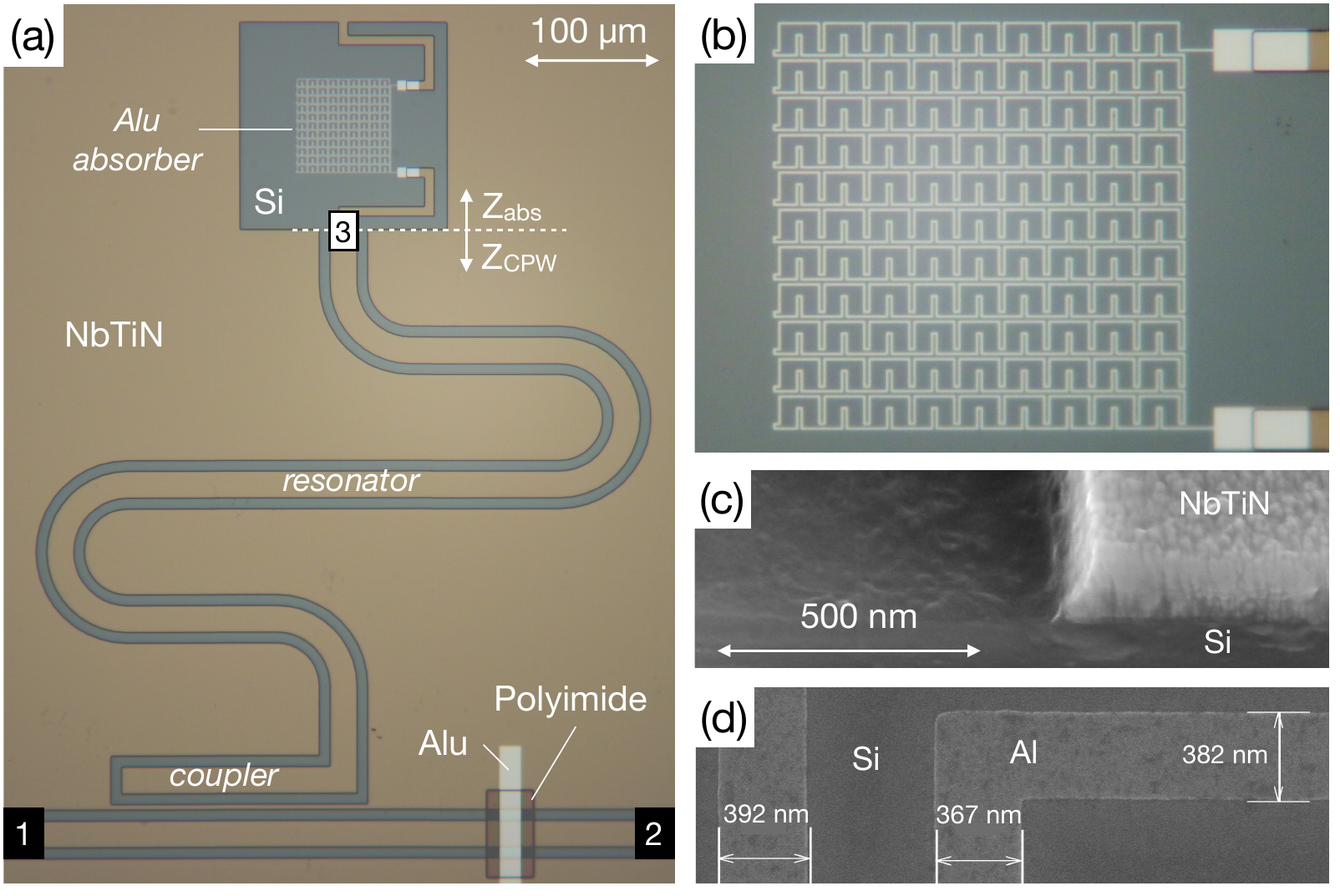}
\caption{(a) Optical micrograph of a single detector, indicating the CPW resonator, coupler, the bridge over the readout line and readout ports 1 and 2. The dash line and arrows indicate the port and de-embedding of the SONNET simulation for the absorber's input impedance. (b) Zoomed in on the absorber structure. (c) Scanning electron microscope (SEM) image of the cross section of the NbTiN, showing the slope obtained during the etching step. This slope guarantees a good step coverage of the Al layer. (d) SEM image of a few lines of the absorber structure indicating the measured line widths.  }
\label{fig_KID} 
\end{figure}

The KIDs consist of the distributed absorbers, which act as both the THz sensitive section and a microwave inductor. The absorber is shorted to the chip ground on one side and connected to the central line of an open ended NbTiN coplanar waveguide (CPW). The device is read-out from port 1 to port 2 as indicated in Fig. \ref{fig_KID}(a). The NbTiN CPW has a (large) line width $S = 20$ $\mathrm{\mu}$m and (large) gap width $W = 8$ $\mathrm{\mu}$m to reduce the Two Level System (TLS) noise of the KIDs \cite{Gao2007}. The CPW length, $l_{CPW}$, is designed to tune the resonant frequency of the KID to the desired values between 2 and 4 GHz. Specifically, we simulated the input impedance of the absorber structure at microwave frequencies, ($Z_{abs} $), using a de-embedded port ‘3’ as indicated in Fig. \ref{fig_KID}(a) in SONNET \cite{ref_SONNET} and the DC resistivity and thickness of the Al. Subsequently, we analytically calculated the input impedance of the open ended NbTiN CPW line, $Z_{CPW} $ from the same port, which is given by the standard expression:

\begin{equation}
\label{Eq_Zin_NbTiN}
Z_{CPW} = Z_{eff, NbTiN}\cdot\cot{\left(\frac{2\pi F}{v_{ph}}l_{CPW}\right)}
\end{equation}

\noindent 
where $Z_{eff, NbTiN}$ is the effective NbTiN line impedance and $v_{ph}$ the phase velocity of the line. The calculation of these parameters is detailed in Appendix \ref{App_CPW}.

\begin{figure}[!t]
\centering
\includegraphics[width=3in]{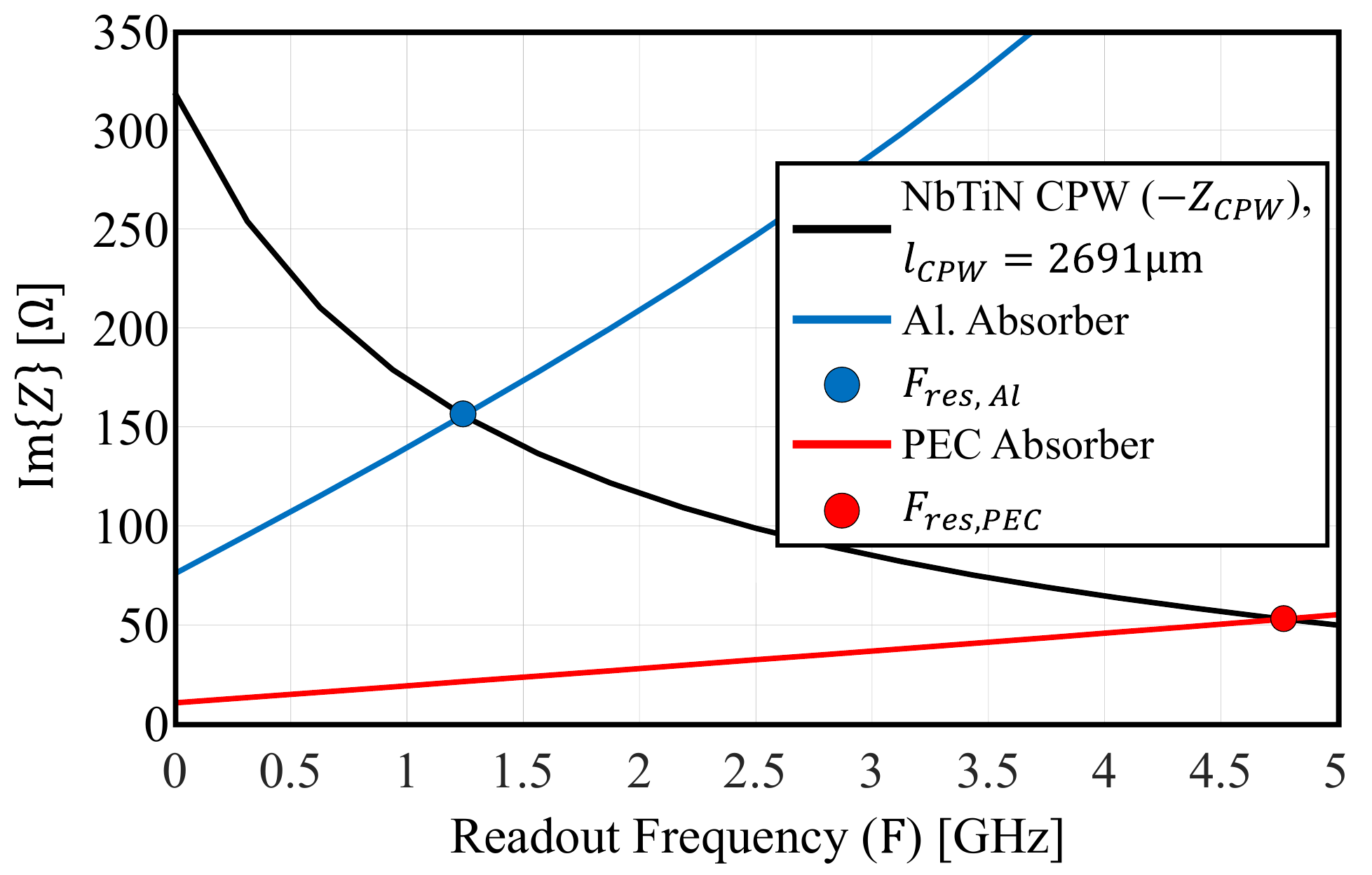}
\caption{Imaginary part of the input impedances of the 2 resonator sections, obtained at port '3' as shown in Fig.\ref{fig_KID}(a). The blue line shows $Z_{abs}$ obtained using SONNET simulations using the measured parameters of the Al film, the red line is obtained by replacing Al with PEC. The black line gives the analytically calculated $Z_{CPW}$ of the CPW line. The intersects give the resonant frequencies, we find $F_{res}=1.94$ GHz for the Al absorber, and $4.80$ GHz for the PEC case.}
\label{fig_Fres} 
\end{figure}

The resonant frequency, $F_{res}$, is given for the condition that the total imaginary impedance of the structure is 0, i.e. $\mathrm{Im}\{Z_{CPW}\} + \mathrm{Im}\{Z_{abs}\}=0 $. As an example we give in Fig. \ref{fig_Fres} the result of this calculation for the lowest frequency KID in our test array, with the longest NbTiN CPW length (see Sec. \ref{sec:IV}). We repeat the calculation for the case where we replaced the Al with a Perfect Electric Conductor (PEC) to obtain $F_{res, PEC}$. From these results we can obtain the kinetic inductance fraction $\alpha_k=1-(F_{\mathrm{res, Al}}/F_{\mathrm{res, PEC}})^2 =0.84$, which is defined as the fraction of the resonator's inductance due to the kinetic inductance of the Al.  Note that these calculations are based on the measured film parameters and geometries, which will be discussed together with the results presented in Section \ref{sec:IVB}. Using this method, we design a KID array with 25 KIDs in a hexagonal packing, with a spacing of $0.75$ mm, with resonance frequencies within a $2.2$-$3.8$ GHz readout frequency range, maintaining a constant fractional frequency difference $\mathrm{d}F_{\mathrm{res}}/F_{\mathrm{res}}= 2.3\cdot 10^{-2} $ between the resonators. Note that the design was done using estimated values of the DC resistance of the Al: a sheet resistance of $ R_{s,est.} \simeq 0.7\Omega/\Box $ and $\mathrm{T_c} = 1.4$ K. 

\section{Experimental Verification}
\label{sec:IV}
In this section, the fabrication, assembly and experimental results corresponding to the lens absorber coupled KID design in Sec. \ref{sec:III} operating at the central frequency of $6.98$ THz are discussed. 

\subsection{Fabrication and Assembly}
\label{sec:IVA}

\begin{figure}[!t]
\centering
\includegraphics[width=3.2in]{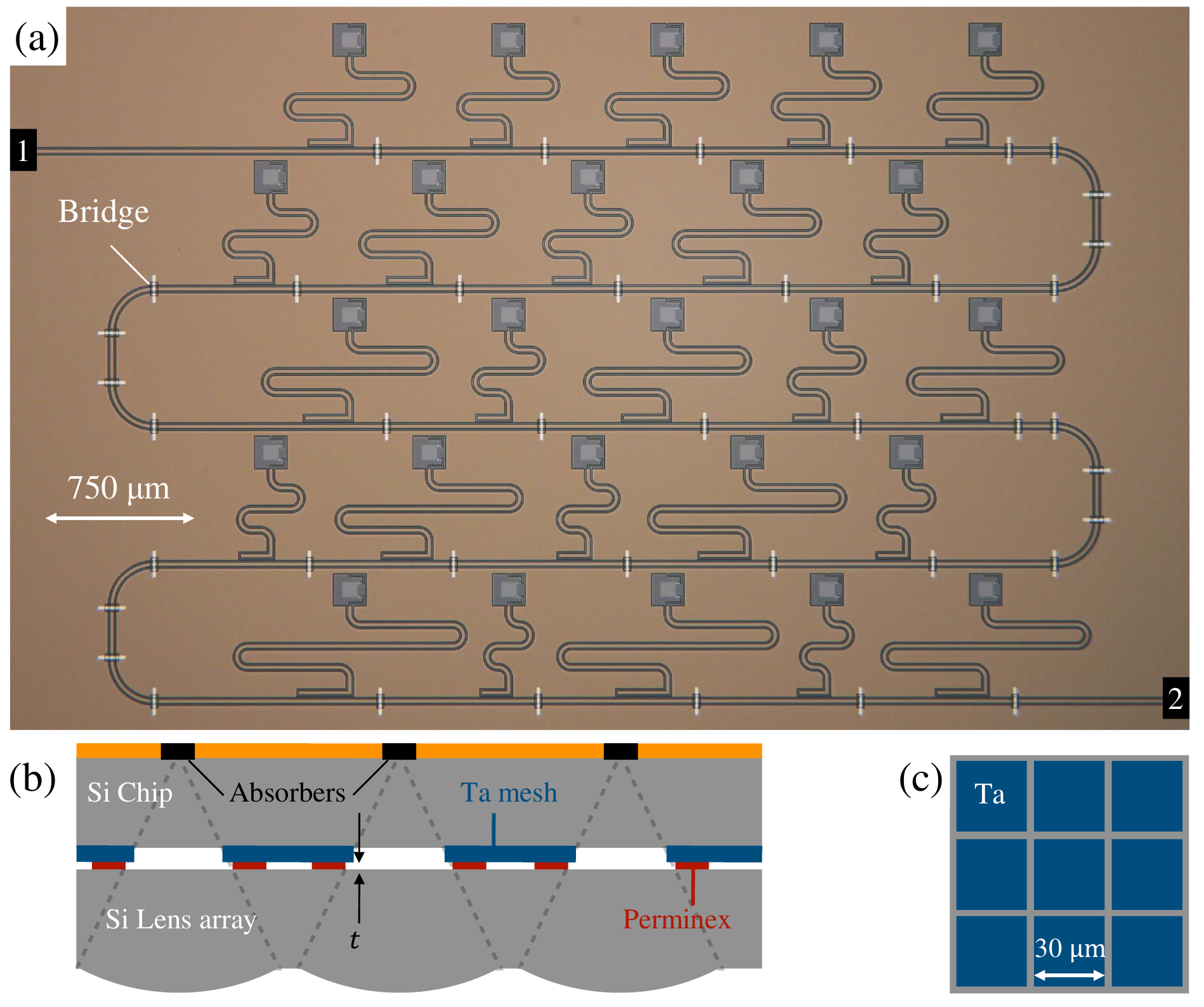}
\caption{(a): Micrograph of the entire KID array, showing $25$ KIDS on a $0.75$ mm pitch in a hexagonal packing. (b) Cross section of the assembled KID-lens array. The dashed lines indicated the radiation between the lenses and the absorbers, where no Ta mesh nor Perminex pillars are placed.  (c) Pattern of the Ta mesh, creating a low-pass transmission making it invisible for signals at the KID readout frequency.  }
\label{fig_Array} 
\end{figure}

We fabricate the devices on a 4 inch Float-Zone $350$$\mathrm{\mu}m$ $R_{\mathrm{s}}\gg$ $10$ k$\Omega$cm $<$$100$$>$ Si wafer. After standard wafer cleaning and a 10 sec. soak in $10\%$ Hydrogen Fluoride (HF), we deposit on the wafer backside a $35$ $\mathrm{nm}$ thick $\beta$-Ta layer with $R_s\;\simeq \;60$ $\Omega/\Box $, which is patterned using contact lithography and dry etching into a capacitive mesh as shown in Fig. \ref{fig_Array}(c). This layer will reduce the impact of cosmic rays \cite{Karatsu2019} as well as the impact of stray radiation propagating through the chip \cite{Yates2017}. Below each absorber location we leave a $300$ $\mathrm{\mu m}$ diameter aperture to pass the beam from the lens array to the absorbers as sketched in Fig. \ref{fig_Array}(b). We subsequently deposit a 200 nm SiO$\mathrm{2}$ sacrificial layer to keep the wafer backside pristine during subsequent processing steps. We then deposit a 150 nm thick NbTiN layer using reactive magnetron sputtering \cite{Thoen2016} which is patterned using contact lithography and dry etching in a SF$_\mathrm{6}-\mathrm{O}_\mathrm{2}$ plasma to create a $\simeq \mathrm{60}^\circ$ sloped edge, shown in Fig. \ref{fig_KID}(c). We measured the sheet resistance of the NbTiN $ \mathrm{R_s}=17.9$ $\Omega/\Box$ and its critical temperature $\mathrm{T_c}$$=15.0$ K using a DC test chip from the same fabrication run and find that both values are identical to the expected ones. Afterwards, we create the support of the bridges over the CPW readout line to balance the ground planes. These bridges are needed to prevent KID-KID crosstalk due to mode conversion on the readout line \cite{Yates2014}.  This is done by spin, expose and cure of the photo-negative Fuji-film\textsuperscript{\textregistered} LTC9305 Polyimide. The last step is the fabrication of the Al absorber, including the contact pads to the NbTiN and the bridges, see Fig. \ref{fig_KID}(a) and (b). We use electron-beam evaporation through a MMA-PMMA (methyl methacrylate - polymethyl methacrylate) bilayer resist mask patterned using a Raith\textsuperscript{\textregistered} EBPG5200 electron beam pattern generator (EBPG) using a dose of $1050$ $\mathrm{\mu}$A/cm$^\mathrm{2}$. The exposed resist is developed using MIBK:IPA (Methyl Isobutyl Ketone - Propanol) 1:3 and etched for 30 sec. in an oxygen plasma to remove residual resist in the opened trenches. We then perform a 30 sec. buffered oxide etch (BOE): water 1:7 for 20 sec. to remove the oxide on the Si and NbTiN and deposit, using e-beam evaporation, a $22$ $\mathrm{nm}$ thick Al layer. We measured on a 4-point DC chip from the same wafer $ R_{s,alu}=$1.21 $\Omega/\Box$, higher than the expected value, and a critical temperature $\mathrm{T_c}$$=1.4$ K, identical to the expected value. The sheet kinetic inductance of the Al is given by $L_s$=1.2 pH calculated using $L_s=hR_s/(\pi\Delta)$, with $\Delta$=1.76$\mathrm{k_b T_c}$ \cite{Leduc2010}. The resist with obsolete Al is removed using a lift-off process in Dimethylformamide (DMF) at $56$$^\circ$C for $1$ hour. This creates, in a single step, the Al absorber, with volume and area of $22.9$ $\mathrm{\mu m^3}$ and $1041$ ${\mathrm{\mu m^2}}$ respectively, as well as the contact to the NbTiN resonator. After this step we remove the sacrificial SiO$\mathrm{2}$ layer on the wafer backside using BOE while protecting the wafer front side with photoresist. This lifts-off any particles or contaminants created during the fabrication.

	To prepare for mounting we apply $500$ nm thick layer of lithographically defined pillars made from Perminex\textsuperscript{\textregistered} 1001 on the chip backside, located only where they will not interfere with the radiation propagating from the lens array to the detector, schematically as shown in Fig. \ref{fig_Array}(b). This creates a $500$ $\mathrm{nm}$ vacuum gap between the lens array and chip, which is thin enough to maintain a good lens-absorber coupling, as shown in the supplementary document.

In Fig. \ref{fig_KID}, we show an optical micrograph of a single detector, and in its panel (d) a SEM image of the absorber lines, which gives an average width of $w_a \simeq 380$ $\mathrm{nm}$, $80$ $\mathrm{nm}$ wider than designed values given in Table \ref{tab_des_para}. These wider lines also reduce the gap widths ($g_1$ and $g_2$) by $80$ $\mathrm{nm}$. The aperture efficiency due to this modified unit cell is shown in Fig. \ref{fig_ap_eff_LensShabs}(a) by dash-dotted lines. In Fig. \ref{fig_Array}(a), we show the centre part of the detector chip under consideration, with $25$ KIDs.

After fabrication, the wafer is diced in $20$ $\mathrm{mm}$ $\times$$20$ $\mathrm{mm}$  chips, with standard optical AZ resist used as protection on the chip backside and PMMA resist on the chip metallisation side. After dicing, we remove the backside protection using PN1000 developer, exposing the Perminex glue pillars. We mount the chip metal side down in an alignment tool, and align the chip to the lens array top, which is fabricated using laser ablation by Veldlaser \cite{Veldlaser} and anti-reflective coated with a $6.9$ $\mathrm{\mu m}$ thick layer of Parylene-C \cite{Ji2000}, as a quarter wavelength matching layer. Note that we align on a marker on the lens array surface, which is displaced in z by $1.07$ $\mathrm{mm}$, from the chip markers, due to the thickness of the lens array.

After alignment, the chip and lens array are pressed together to create the required pressure of $0.58$ $\mathrm{MPa}$ on the Perminex. We then bake the assembly in a pre-heated oven at $180$$^\circ$C for $15$ minutes. In this step the chip and lens array reach a temperature exceeding $150^\circ$C for 5 minutes, which was evaluated in a test using thermal stickers. This creates a permanent bond between the chip and lens array. We remove the remaining front side PMMA using acetone.

\subsection{Measured Performance}
\label{sec:IVB} 

We mount the chip-lens array in a suitable holder, which is light tight, inside a completely light-tight box thermally anchored to the cold finger of an Adiabatic Demagnetisation Cooler (ADR) \cite{Baselmans2012} at $130$ mK. To illuminate the detector array we use a thermal black body radiation source coupled to the $3$ K stage of the cryogenic system which can be heated up to $40$ K. The black body, sample holder and light tight box are equipped with a set of commercial metal mesh filters, consisting of two band pass filters (BPF) and two high pass filters (HPF) from Celtic Terahertz Ltd. defining a pass-band around $6.98$ THz with extreme rejection at longer wavelengths. The total throughput from the radiator to the detector chip is defined by a $15$ $\mathrm{mm}$ diameter aperture at $55$ $\mathrm{mm}$ from the chip metallisation side at the exit of the light tight box, which defines a full width opening angle of $15.95$$^\circ$. For a detailed description of this setup and its cross section, we refer to Appendix \ref{App_setup} and Fig. \ref{fig_SetupA}(a), respectively. 

\begin{figure}[!t]
\centering
\includegraphics[width=3.5in]{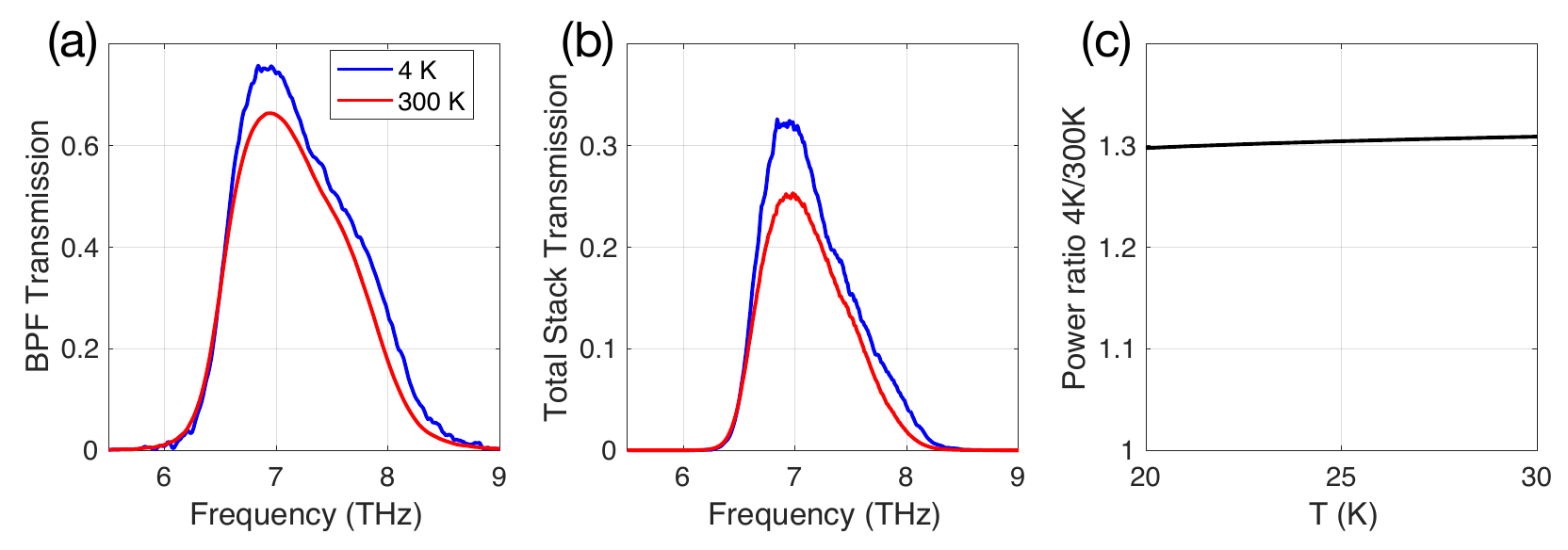}
\caption{(a) Room temperature and $4$ K transmission for one BPF of our setup. (b) Room temperature and $4$K transmission for the entire filter stack of 2 BPF's and 2 HPF's. (c) Ratio of the power coupled to the detector for the 4K data divided by the 300K data as function of black body temperature within the $\mathrm{T_{bb}}$ temperature range of $20-30$ K where we measure the optical efficiency. }
\label{fig_FiltersA} 
\end{figure}

A key issue with the filter stack is that its in-band transmission is low, $25\%$, using the data provided by Celtic THz Ltd, which is obtained at $300$ K. In our experiments we use the filters at $4$ K or $130$ mK, where the dielectric and metallic losses will be lower than at 300K. Hence, the in-band transmission will be higher than based upon the $300$ K data, which can lead to an overestimation of the coupling efficiency, as found in Ref. \cite{Baselmans2022}. To eliminate as much as possible such a systematic error we measured a single band pass filter of our filter stack  in a commercial FTS at $4$ K, the result is shown in Fig. \ref{fig_FiltersA}(a). We observe a clear increase in transmission for the $4$ K data, as well as a significant broadening of the filter bandwidth, coming from an upshift of the high frequency edge. For the two HPF's in our setup we relied upon data from Celtic THz for a similar filter as the one we used. These measurements show only a $2.5\%$ increase in transmission when going from $300$ K to $4$ K. In Fig. \ref{fig_FiltersA}(b), we show the resulting frequency dependent transmission of the \emph{entire} filter stack (2 BPF's and 2 HPF's) at $4$ K, together with the $300$ K data as reference. In panel (c) we show the  relative increase of the total transmission in the setup integrated over frequency for the black body temperature range where we will obtain the detector coupling efficiency. We observe an increase in power of $30\%$ for the $4$ K data with respect to the original $300$ K filter data. It is noteworthy that this is only caused by the peak transmission increase, the upshift of the high frequency edge has minimal effect as the Planck brightness of the radiator decreases exponentially for all black body temperatures plotted and the power transmitted is dominated by the low frequency edge and the peak filter stack transmission.

\begin{figure}[!t]
\centering
\includegraphics[width=3.5in]{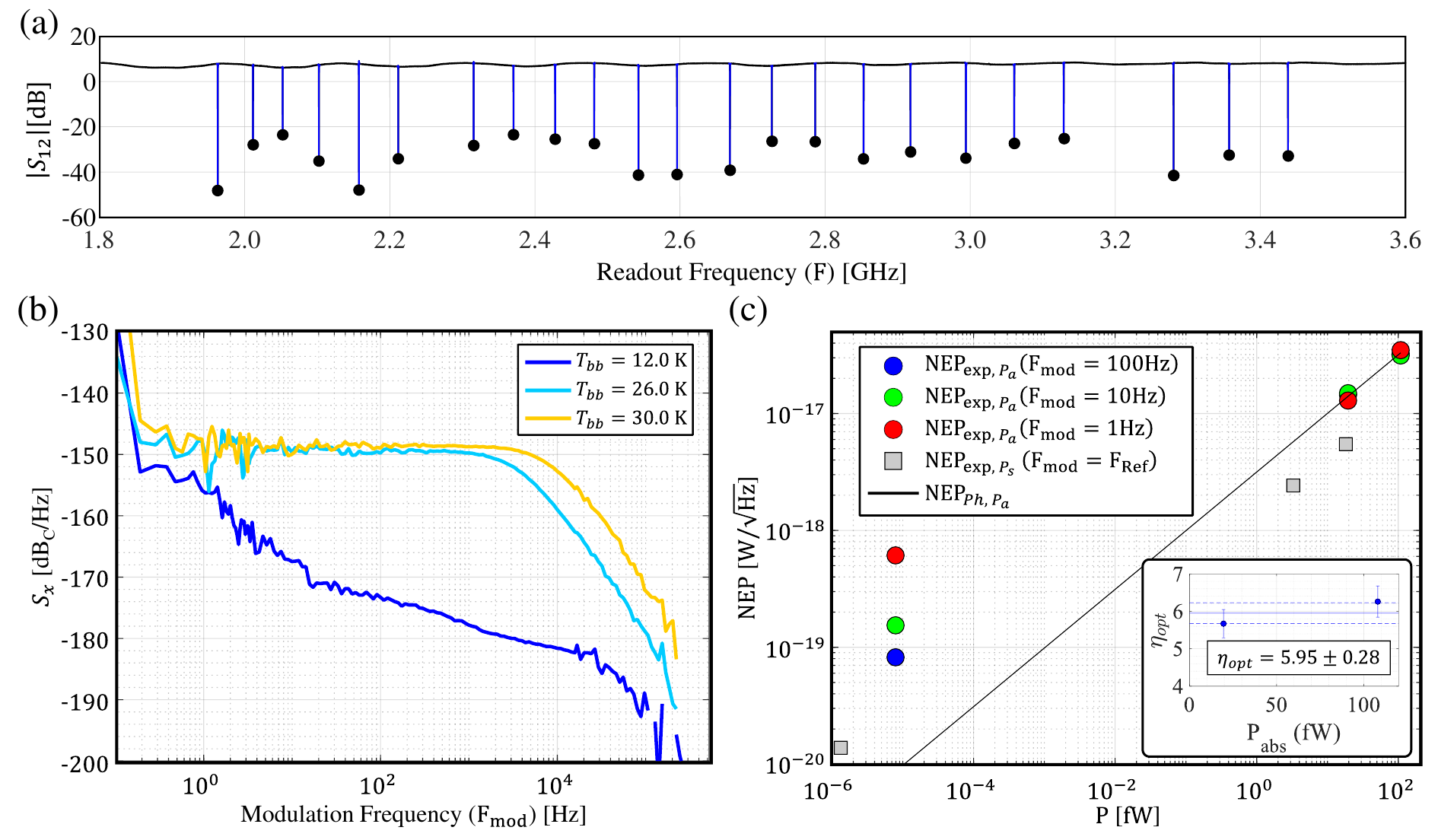}
\caption{ (a) Frequency sweep at $130$ mK and negligible radiator power showing $23$ out of $25$ KIDs. (b) Noise spectra for KID $10$ $F_{\mathrm{res}}=2.48$ GHz for the three radiator temperatures used in this experiment. (c) Noise equivalent power as a function of absorbed power for KID $10$, with in the inset the measured optical efficiency for the two higher radiator temperatures. }
\label{fig_Result} 
\end{figure}

\begin{figure}[!t]
\centering
\includegraphics[width=3.5in]{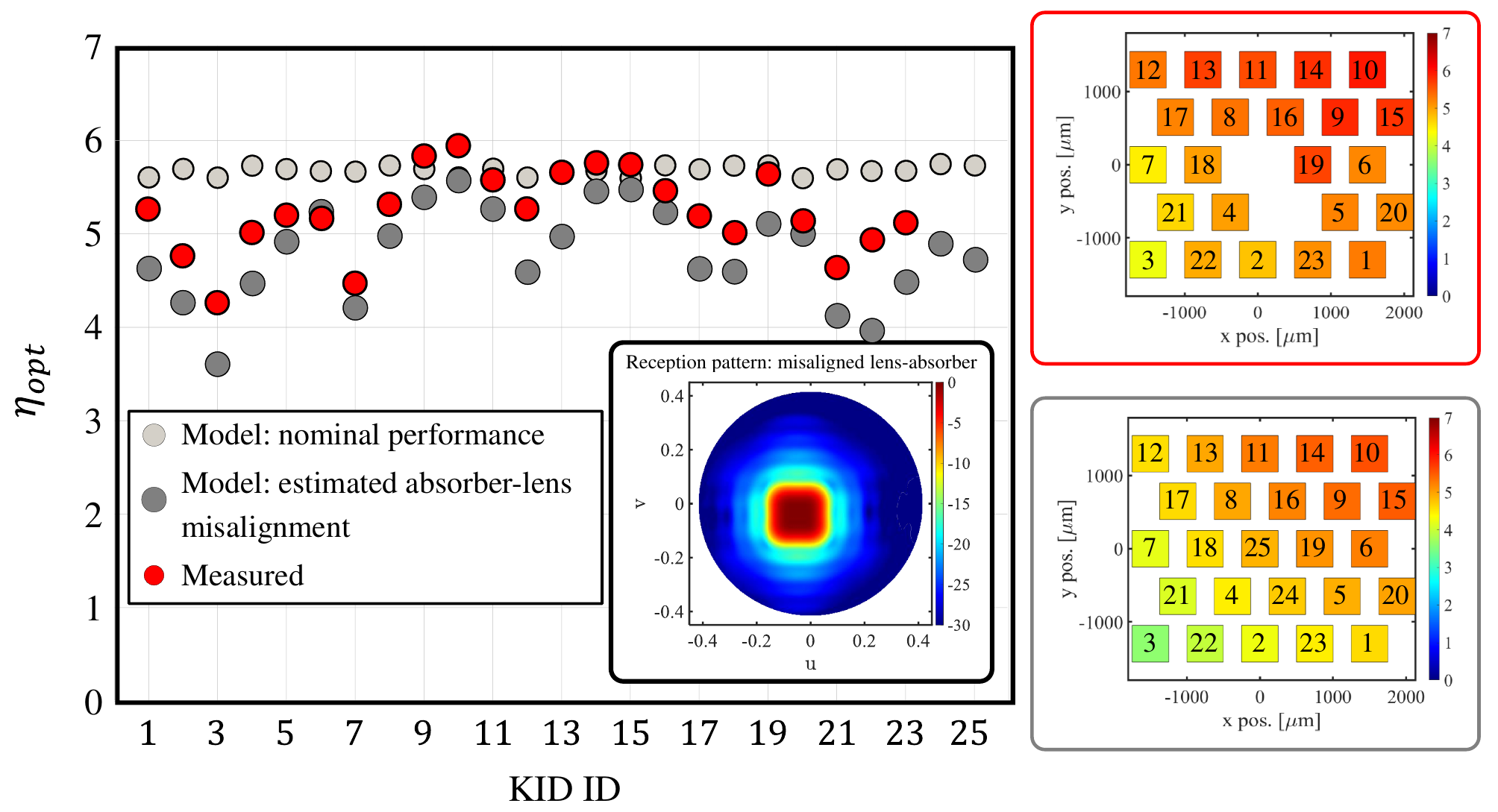}
\caption{Optical efficiency for all KIDs, the inner inset shows the power pattern of the lens absorber, at the centre of the array, when the absorber is laterally displaced with respect to the lens centre which corresponds to the case of the assembled devices. The outer top and bottom insets indicate the spatial variation of the measured and modelled optical efficiency for each KID, respectively, where the KID's IDs are also provided.} 
\label{fig_Result_eta_opt_15} 
\end{figure}

In the first experiment we measured the resonant frequency of all the KIDs at the bath temperature of $\mathrm{T_{Bath}}=130\;$ $\mathrm{mK}$ and a radiator temperature of $\mathrm{T_{BB}} = 2.7$ K. At this temperature the power radiated in-band is negligible ($\leq \mathrm{10^{-23}}$ $\mathrm{W}$, as shown in Fig. \ref{fig_SetupA}(c) in Appendix \ref{App_setup}). As can be seen from Fig. \ref{fig_Result}(a), we find $23$ out of the $25$ KIDs, which represent a yield of $92\%$, with a coupling Q factor $\mathrm{Q_c}\;=\;2.1\pm0.6\cdot10^4$ and internal Q factor $\mathrm{Q_i}\;=\;8.2\pm0.5\cdot10^6$. We also find that the fractional frequency spacing of the KIDs  $\mathrm{d}F_{\mathrm{res}}/F_{\mathrm{res}}=0.023\pm0.0019$ as designed. The standard deviation of the fractional frequency scatter, $\sigma_{\mathrm{d}F_{\mathrm{res}}/F_{\mathrm{res}}}\mathrm{=1.9\cdot10^{-3}}$ is an order of magnitude worse than the value reported by \cite{Shu2018} of $\sigma_{\mathrm{d}F_{\mathrm{res}}/F_{\mathrm{res}}}\mathrm{=1.8\cdot10^{-4}}$, but obtained without post fabrication corrections. The lowest frequency resonator has $F_{\mathrm{res}}$ $= 1.96$ GHz, very close to the calculated value based upon the measured film parameters and device geometry of 1.94 GHz which is shown in Fig. \ref{fig_Fres}.

In the second experiment, we measured the NEP as well as the optical efficiency for all KIDs, in an identical manner as in \cite{Baselmans2022}. The definition and derivation of this optical efficiency for a multi-mode detector is provided in Appendix D. We first obtain the experimental NEP as a function of the source power: 

\begin{equation}
\label{Eq_NEP_Ps}
\mathrm{NEP}_{\mathrm{exp},P_s}=\sqrt{S_{\theta} }  \left(\frac{d \theta}{\mathrm{d}P_s} \right)^{-1} 
\end{equation}

\noindent
Here $S_{\theta}$ is the KID phase noise, obtained from $128$ seconds of time domain data taken at a constant radiator temperature, $d\theta$ is the KID phase response to a change in the radiator power $\mathrm{d}P_s$. This was obtained by measuring the KID phase while performing a small temperature sweep of the radiator and performing a linear fit from the measured data. In all cases $P_s$ is calculated from the radiator temperature and the filter transmission as discussed in Appendix \ref{App_setup}. 

In Fig. \ref{fig_Result}(b) we show the reduced frequency noise $S_x=S_{\theta}/(4Q_i)^2$ of KID 10 ($F_{\mathrm{res}} =2.79$ GHz) for three different radiator temperatures. The noise at $\mathrm{T_{BB}}=12$ K, $P_{abs}=$ 0.6 aW, which represents still fully dark conditions, has a strong 1/f component. At the two higher black body temperatures the noise is white with a roll-off given by the quasiparticle lifetime. In this regime, the KIDs are photon noise limited \cite{Janssen2013}, with a sensitivity dominated by the photon arrival rate fluctuations. Now we can use Eq. (\ref{Eq_eta_m}) in Appendix \ref{App_opt_coup} to obtain the optical efficiency with respect to a single mode and one polarisation $\eta_{opt, f_0}$ from $\mathrm{NEP}_{\mathrm{exp},P_s}$ which is shown by the grey coloured squares in Fig. \ref{fig_Result}(c). The inset indicates the optical efficiency found using this method. We obtain an average value of $\eta_{opt}=5.95$, which corresponds to receiving $5.95$ times the power of a single mode at a single polarisation. We repeat this procedure for all $23$ KIDs and give the complete data set in Fig. \ref{fig_Result_eta_opt_15}. We observe an optical efficiency that scatters between $4.3-5.95$. The maximum coupling is in good agreement with the calculated value using the measured absorber geometry and Al resistivity, and corresponds to an aperture efficiency of  $54\%$ as shown in Fig. \ref{fig_ap_eff_LensShabs}(a). 
We attribute the scatter in efficiency to small lens-absorber misalignments due to a systematic shift between lens array and chip, which was caused by an $\approx$ $1.3$$^\circ$ off-vertical alignment between the chip plane and microscope z-axis used to align the lens array to the chip. We correct for this using an independent measurement of the coupling efficiency using a small aperture in our setup as discussed in Appendix \ref{App_Result}, and shown in Fig. \ref{fig_Result_opt_4.5mm}. We find a misalignment of the lens array that can be described by a shift common to all detectors of  $\simeq 29.9\hat{x} + 15.1\hat{y}$ $\mathrm{\mu m}$ with respect to a lens focal point, which is consistent with a separate measurement of the angle between the microscope z-axis and the chip plane. As a result, a corresponding beam tilt is present in the reception power pattern of the fabricated lens absorbers. The resulting tilted reception power pattern of the lens absorbers due to this misalignment is shown in the inset of Fig. \ref{fig_Result_eta_opt_15} which repoints the beam maximum value (with respect to a lens optical axis) to $4.6^{\circ}$ colatitude and $207^{\circ}$ azimuth angles. Moreover, we expect slight rotation misalignment between the lens array and absorbers as well as slight misalignment variation for each of the 25 absorber-lenses. These higher order misalignment suspicions are not considered in the corrected model. Despite the described misalignment complications, we observe a good agreement between the measured and calculated results for both aperture sizes. It is noteworthy that the misalignment correction does not change the maximum coupling for the best detectors. 

The agreement between measured and estimated optical efficiency implies that the experimental aperture efficiency of the lens-absorber is also in agreement with the one estimated by the model, i.e. averaged in polarisation of $\eta_{ap}\mathrm{(6.98\;THz)=0.54}$ [see the grey coloured circle in Fig. \ref{fig_ap_eff_LensShabs}(a)]. The reasoning for the validity of this indirect verification is as follows: The optical efficiency depends both on the reception power pattern (i.e. focusing efficiency) as well as the aperture efficiency, as described in Appendix D, Eq. (\ref{Eq_eta_op}). However, a negligible ambiguity is expected on the lens-absorber's reception power pattern due the two following: i) As shown in Fig. \ref{fig_lens_absorber_ideal_PRx} in Appendix \ref{App_fig_mert}, the reception power pattern depends strongly on the absorber total size and the lens diameter, but not the details of a unit cell nor the stratification. In the described prototype, the physical size of the absorbers and lens diameters are well controlled. ii) In the experiment with the smaller radiator aperture (see Fig. \ref{fig_Result_opt_4.5mm} in Appendix \ref{App_Result}), the optical efficiency depends strongly on the spillover efficiency of the setup (see Eq. (\ref{Eq_SO_eff_setup})) and in turn to the reception power pattern of the lens absorber. In this case, the measured optical efficiency is still in fair agreement with the one from the model. Therefore, it is concluded that optical efficiency verification indirectly leads to an experimental validation of the aperture efficiency.  

After obtaining the optical efficiency, we can obtain $\mathrm{NEP}_{\mathrm{exp}, P_{\mathrm{abs}}}$, which is shown in Fig. \ref{fig_Result}(c) for KID 10 using Eq. (\ref{Eq_NEPph_abs}) in Appendix D. We reach a limiting NEP of $\mathrm{NEP}_{\mathrm{exp}, P_a}=8\cdot10^{-20}\mathrm{W\sqrt{Hz}}$ at a modulation frequency, $F_{\mathrm{mod}}$, of $100$ Hz, which increases at  lower modulation frequencies due to the 1/f noise. This result is typical for all devices, as is shown in Fig. \ref{fig_ResultA1}  in Appendix \ref{App_Result}. We also refer to Appendix \ref{App_Result} for the noise and NEP data at wider range of radiator temperatures (Fig. \ref{fig_ResultA2}), to illustrate the radiation power dependence of the NEP and the noise of the detectors presented in this work. This noise transforms from an 1/f spectrum at low absorbed powers to a white spectrum for $P_{abs}>\mathrm{69\;fW}$. It is interesting to compare these results to recent work from Day \emph{et al.} \cite{Day2024}, which discusses a similar detector but optimised for higher frequencies. These devices have a lower NEP than the devices presented here, especially at low modulation frequencies, and a shorter maximum quasiparticle recombination time of 0.5 msec. This can possibly be attributed to one or more of the following reasons: i) differences in the fabrication methodology, ii) the absorber volume, which is $4$ fold larger in our devices and iii) the fact that the quasiparticle density at low temperature and negligible power case can be limited by the readout power, which is both absorber volume and readout frequency dependent \cite{deRooij2025}. 


\section{Conclusion}

In this paper a detector concept based on lens absorber coupled KIDs is proposed to address the requirements of future space-based FIR astronomical instruments in terms background limited sensitivity and FPA scalability. The EM coupling of the detectors placed below focusing lens components is modelled via a spectral technique developed in previous contributions. This technique provided an accurate and numerically efficient manner to describe the relevant figures of merit, design guidelines, and tolerance studies for lens absorbers. Two dual polarised lens absorber couple KIDs operating at central frequencies of $6.98$ and $12$ THz, relevant to PRIMAger bands, were presented. The design for $6.98$ THz was fabricated as a hexagonal array of 25 detectors with a yield of $92\%$. The performance of the prototype array in terms of optical coupling was experimentally validated against the one estimated by the model with good agreement. This also indirectly validated the aperture efficiency of the lens absorbers against the expected value of  $\simeq 54\%$.  The NEP of the absorber based KIDs was also quantified with a minimum achieved value of $8\times 10^{-20}$ $\mathrm{W/\sqrt{Hz}}$ satisfying the targeted sensitivity requirements. The experimental verifications shown in this work indicate that the design, fabrication, and performance of the proposed lens absorber coupled KIDs is well understood providing confidence in developing flight detectors for space-based FIR astronomical instruments, and in particular PRIMA.          



{\appendices
\section{EM Figures of Merit for a Lens Absorber }
\label{App_fig_mert}

In this Appendix, the figures of merit, derived in \cite{Llombart2018} for bare absorbers below reflectors, for describing the EM performance of a lens absorber are reviewed.

\textbf{Aperture efficiency}, $\eta_{ap}$, indicates how much of the incident point source power is captured by the absorber \cite{Llombart2018}:

\begin{equation}
\label{Eq_eta_ap}
 \eta_{ap}(f) = \frac{ \max{ \{P_{\mathrm{abs}}(f,\theta_{l}^{\mathrm{pw}},\phi_{l}^{\mathrm{pw}}) }\}}{P_{\mathrm{inc}}}
\end{equation}

\noindent
where $P_{\mathrm{inc}}=1/(2\zeta_0)|E_{0}^{\mathrm{pw}}|^2A_{\mathrm{lens}}$ is the  power illuminating the lens surface via an incident plane wave, $\zeta_0$ is the characteristic impedance in free space, $E_{0}^{\mathrm{pw}}$ is the amplitude of the incident plane wave, and $A_{\mathrm{lens}}$ is the area of the lens aperture. 

\textbf{Reception power pattern} of the lens absorber, indicates its normalised angular response to point sources \cite{Llombart2018}:

\begin{equation}
\label{Eq_PRx}
 F(f,\theta_{l}^{\mathrm{pw}},\phi_{l}^{\mathrm{pw}}) = \frac{ P_{\mathrm{abs}}(f,\theta_{l}^{\mathrm{pw}},\phi_{l}^{\mathrm{pw}}) }{\max{ \{P_{\mathrm{abs}}(f,\theta_{l}^{\mathrm{pw}},\phi_{l}^{\mathrm{pw}}) }\}}
\end{equation}

 The normalised reception power pattern for two absorber sampling sizes and the two ideal absorbers, described in Sec. \ref{sec:IIB}, is shown in Fig. \ref{fig_lens_absorber_ideal_PRx} and compared to the diffraction limited case, i.e. to their corresponding Airy patterns.

\begin{figure}[!t]
\centering
\includegraphics[width=3in]{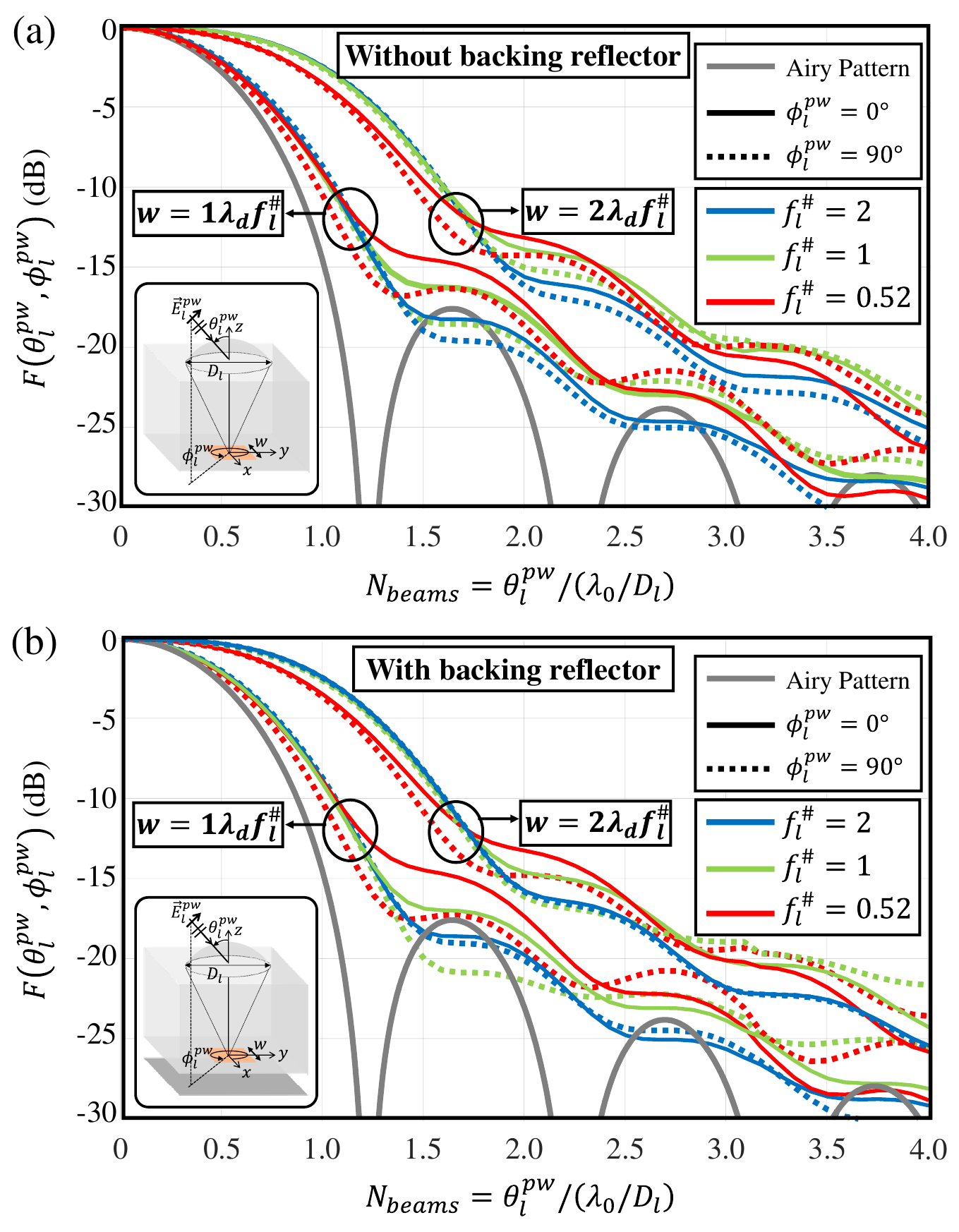}
\caption{The normalised reception power pattern for two absorber sampling rates and three lens f-number cases: ideal absorber (a) without, and (b) with quarter wavelength backing reflector. The insets indicates the scenarios under consideration. The incident plane wave illuminating the lenses is y-polarised.  }
\label{fig_lens_absorber_ideal_PRx} 
\end{figure}

\textbf{Focusing efficiency} compares the angular selectivity of the lens absorber against the diffraction limit \cite{Llombart2018}:

\begin{equation}
\label{Eq_eta_f}
\eta_f(f)= \dfrac{\lambda_0^2/A_{\mathrm{lens}}}{\int_{0}^{2\pi}{\int_{0}^{\pi/2}{F(f,\theta_{l}^{\mathrm{pw}},\phi_{l}^{\mathrm{pw}}) \sin{\theta_{l}^{\mathrm{pw}} \mathrm{d}\theta_{l}^{\mathrm{pw}}\mathrm{d}\phi_{l}^{\mathrm{pw}} } } } }
\end{equation}

\noindent
where $\lambda_0$ is the free space wavelength at frequency $f$; the numerator and the denominator are the solid angles in diffraction limit \cite{Goodman2015}, and the one of the lens absorber, respectively.

\section{Additional EM results for the designed lens absorbers  }
\label{App_EM}

In this appendix, additional results related to the designed lens absorbers operating at the central  frequencies of $6.98$ and $12$ THz are reported.

Lens absorbers performance in terms of the reception power pattern over their operation frequency bands is shown in Fig. \ref{fig_PRx_LensShabs}. These power patterns are symmetric over the azimuth direction with expected frequency variations due to the size of the absorbers in terms of wavelength. The  spectral response of the absorbers to broadside TE and TM plane wave incidents in Si as a function of operation frequency is shown in Fig. \ref{fig_PW_Shabs_vs_freq} while the 2D spectral response of the $6.98$ THz absorber as a function of incident angles at the central frequency is shown in Fig. \ref{fig_PW_Shabs_vs_angbc}.

\begin{figure}[!t]
\centering
\includegraphics[width=3in]{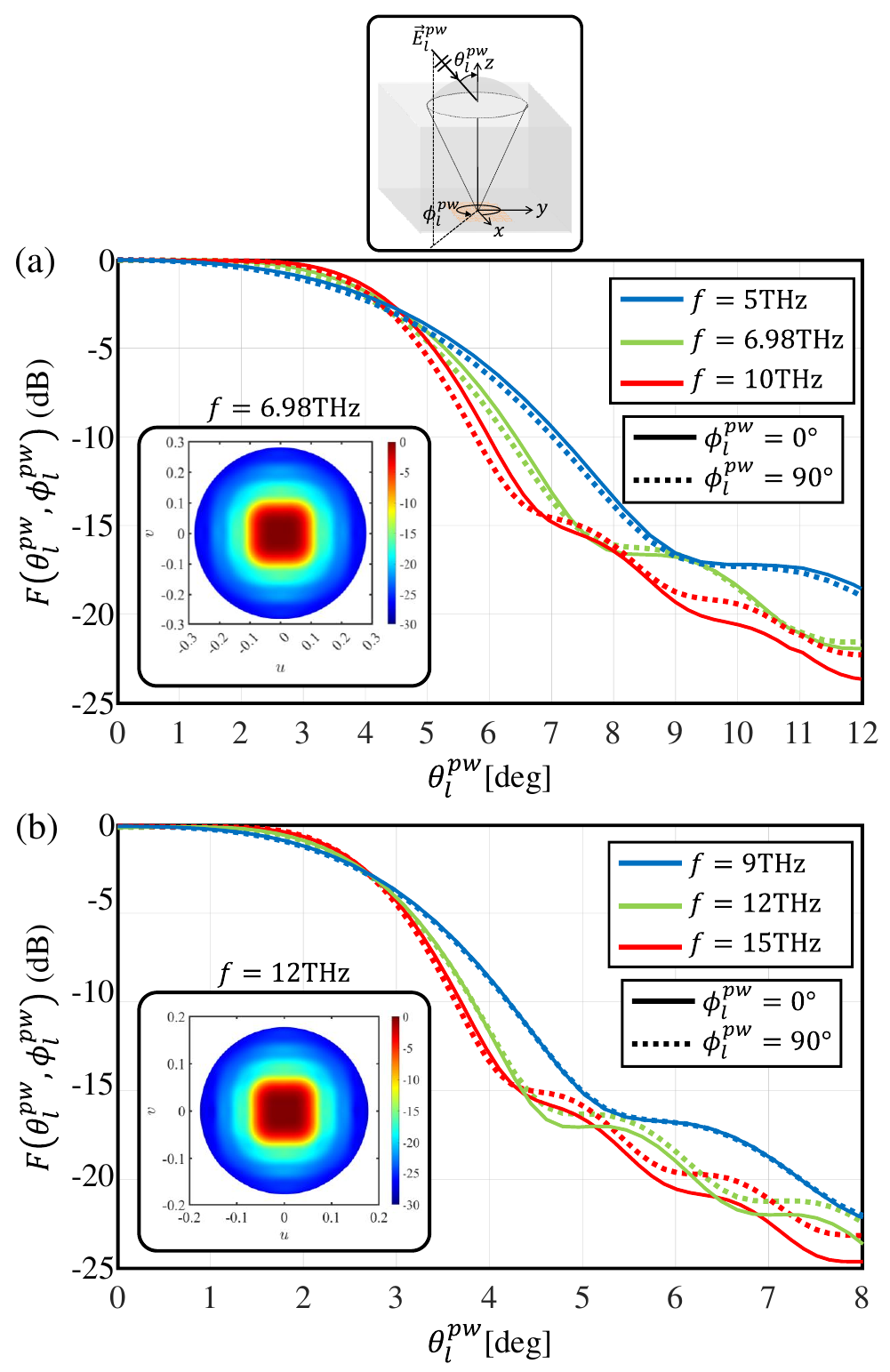}
\caption{The reception power pattern of the designed (a) $6.98$ and (b) $12$ THz lens absorbers against the incident angle of the plane wave illuminating the lens for the lower, centre, and higher frequency points of their operation band. The patterns are the averaged response of the detectors to a combination of the two linear polarisations.   }
\label{fig_PRx_LensShabs} 
\end{figure}

\begin{figure}[!t]
\centering
\includegraphics[width=3in]{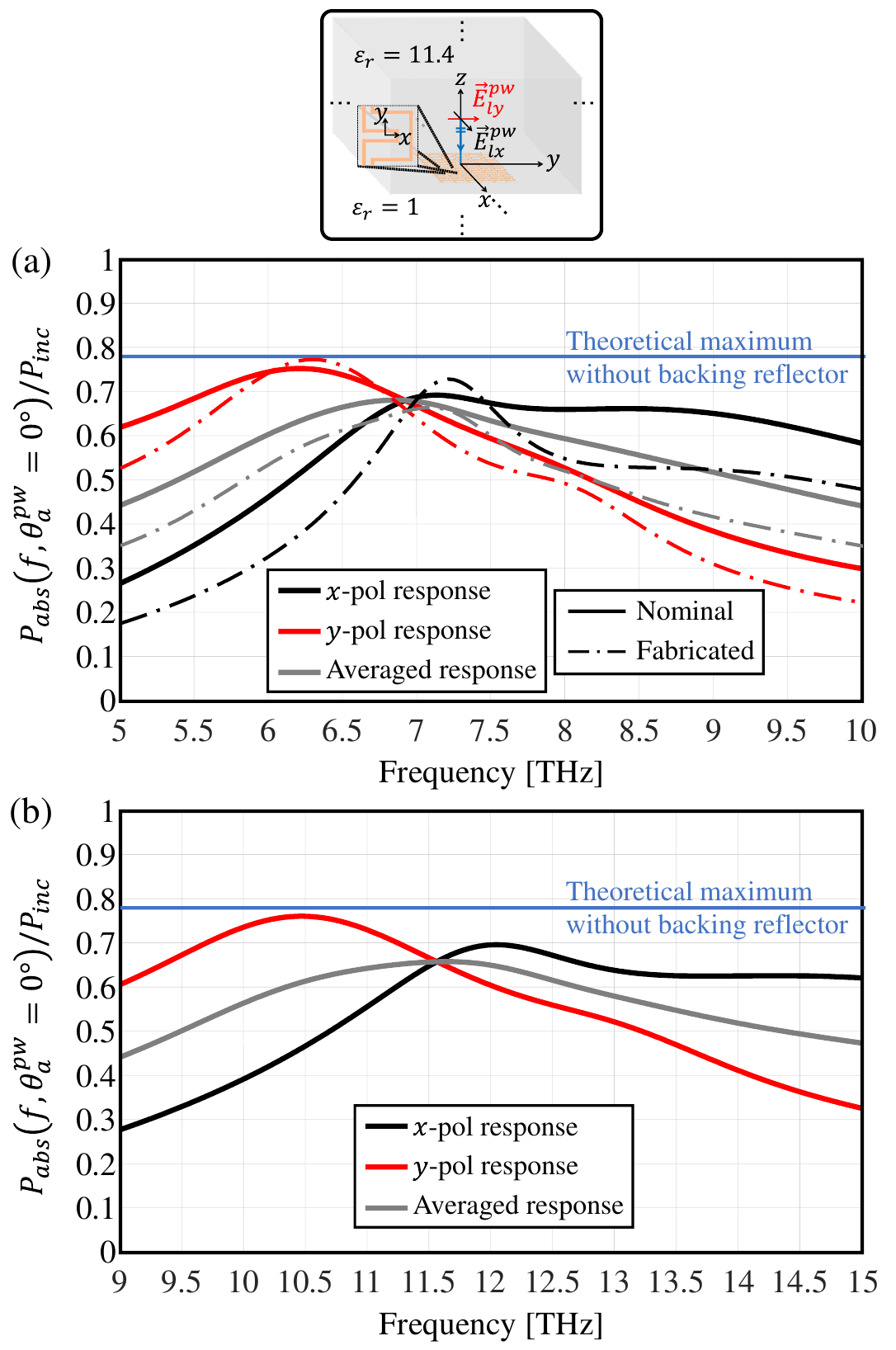}
\caption{ The broadside plane wave response of (a) $6.98$ THz absorber, and (b) $12$ THz absorber as a function of operation frequency. The insets indicate the orientation of the unit cells with respect to $x$- and $y$-polarised plane waves. The solid and dotted lines correspond to the designed and fabricated devices, respectively. The blue solid lines indicate the theoretical maximum value based on Fig. \ref{fig_PW_id_abs}.  }
\label{fig_PW_Shabs_vs_freq} 
 \end{figure}

\begin{figure}[!t]
\centering
\includegraphics[width=3in]{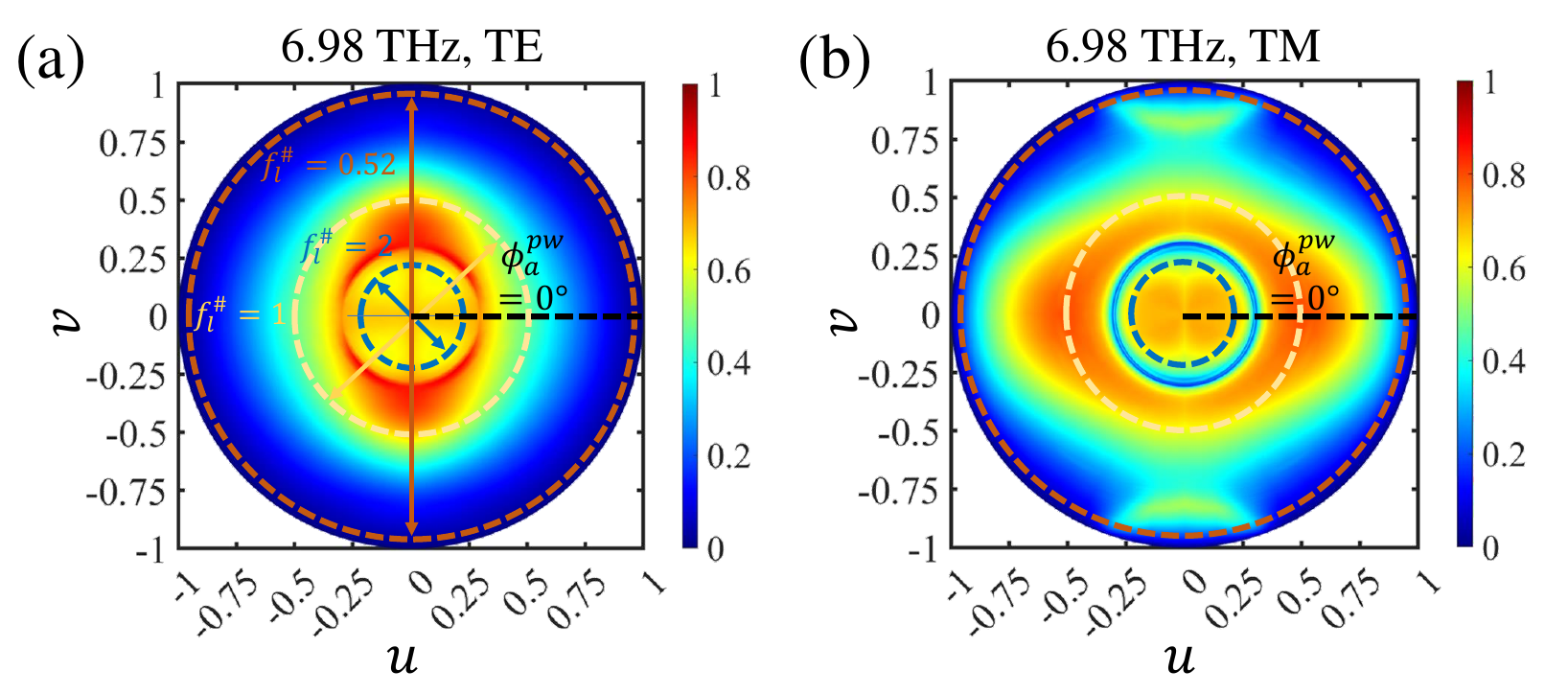}
\caption{ (a) TE and (b) TM polarised plane wave responses of $6.98$ THz absorber projected into $u$-$v$ coordinates, where $u=\sin{\theta_a^{\mathrm{pw}}}\cos{\phi_a^{\mathrm{pw}}}$ and $v =\sin{\theta_a^{\mathrm{pw}}}\sin{\phi_a^{\mathrm{pw}}}$. The indicated angular regions correspond to the three considered lens f-number cases, and data of the indicated cut is shown in Fig. \ref{fig_PW_Shabs_vs_ang}.  }
\label{fig_PW_Shabs_vs_angbc} 
\end{figure}

\section{Calculation of the characteristic impedance of a NbTiN CPW line}
\label{App_CPW}

The characteristic impedance of the NbTiN CPW line, $Z_{eff, NbTiN}=\sqrt{L_{tot}/C_{geo}}$, can be calculated by including the NbTiN kinetic inductance $L_k$ at the readout frequency $F$ in the total inductance, $L_{tot}=L_{geo}+L_k$, where $L_{geo}$ is the geometric inductance.
The sheet kinetic inductance $L_s$ can be calculated at $hF,k_BT<<2\Delta$ using \cite{Leduc2010}:
 \begin{equation}
\label{Eq_Ls}
L_s = \frac{hR_s}{\pi \Delta}
\end{equation}
with $R_s$ the NbTiN sheet resistance, $\Delta=1.76\cdot k_BT_c$ the superconducting energy gap, $k_B$ the Boltzmann constant, and $h$ Planck's constant. The kinetic inductance per unit length of the CPW line can then be obtained by following Collin \emph{et al.} \cite{Collin1992}:

\begin{align}
\label{Eq_Lk}
&L_k = L_s\cdot(g_g+g_c)\;\;with\\
\begin{split}
&g_{c}(S,W) = \\
 &\frac{1}{4S(1-k^2)K^2(k)}\left[\pi + \ln\left(\frac{4\pi S}{d} \right) -k\ln\left( \frac{1+k}{1-k} \right) \right] 
 \end{split}\\
 \begin{split}
&g_{g}(S,W) = \\
& \frac{k}{4S(1-k^2)K^2(k)}\left[\pi + \ln\left(\frac{4\pi (S+2W)}{d} \right) -\frac{1}{k}\ln\left( \frac{1+k}{1-k} \right) \right] \\
\end{split}\\
&with\;k=\frac{S}{S+2W}
\end{align}

Here $S$, $W$ are the NbTiN CPW line and gap widths respectively, $d$ is the metal thickness and $K$ is the complete elliptical integral of the first kind. The phase velocity can now be calculated using the standard expression $v_p=\sqrt{1/(L_{tot}C_{geo})}$.

\section{Experimental Setup}
\label{App_setup}

\begin{figure}[!t]
\centering
\includegraphics[width=3in]{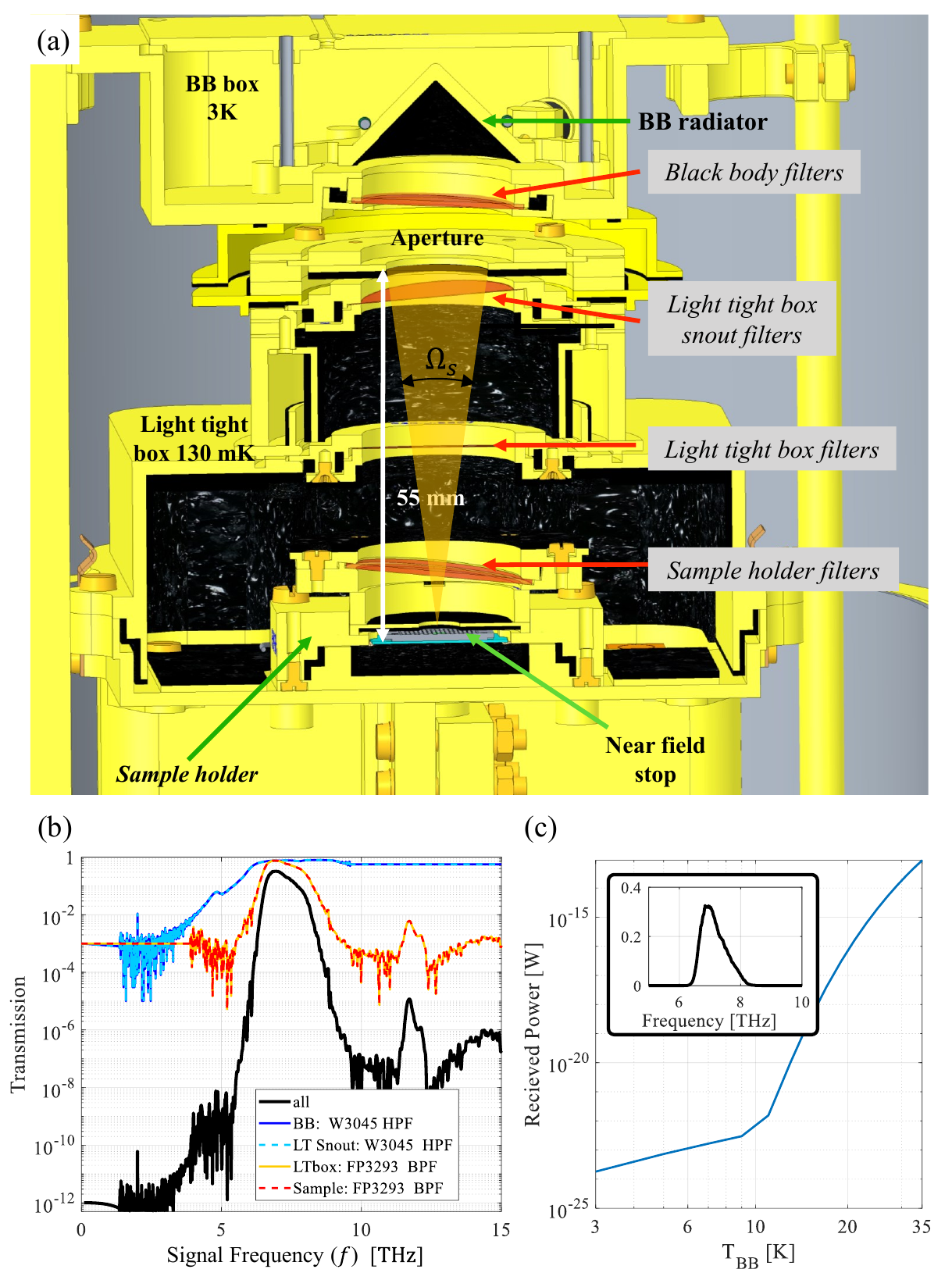}
\caption{(a) Cross sectional view of the experimental setup. (b) Transmissions of the individual filters in the setup, together with the total transmission. (c) Source power as function of radiator temperature for a single mode (and one polarisation). The inset shows the total filter transmission, same as in panel (b), but in a linear scale.  }
\label{fig_SetupA} 
\end{figure}

We give a cross section of the experimental setup in Fig. \ref{fig_SetupA}. It is conceptually similar to the one originally presented in  \cite{Baselmans2012} and used also in \cite{Baselmans2022}. We use a box-in-box at $130$ mK consisting of a light tight box, with all metal slits equipped with stray light labyrinths and their interior coated over by large area absorbers. The absorber material is Epotek920 epoxy with $3\%$ by weight carbon black with SiC grains. We first apply a thin layer of epoxy, then sprinkle the SiC grains, shake off the loose grains and bake out at 90 degree for half an hour. After that we cover this layer with a thin layer of the Epotek/carbon mix, making sure the surface roughness is maintained. For large areas we use $1$ $\mathrm{mm}$ SiC grains, for the labyrinths we use $0.2$ $\mathrm{mm}$ SiC grains. We modified the setup with respect to recent work in \cite{Baselmans2009} with respect to the clamping of the filters: every filter clamp is now carefully equipped with radiation absorbers, and we added a second filter stage on the light tight box. Both additions were needed to prevent any detector response below a radiator temperature of $12$ K, where the in-band power is $\approx10^{-21}$ $\mathrm{W}$, corresponding to a photon rate of around $0.1/\mathrm{sec}$. Note that all filter stages in Fig. \ref{fig_SetupA}(a) are slightly angled to prevent standing waves. In Fig. \ref{fig_SetupA}(b) we show the filters used: high pass filters at the radiator and snout position, and band pass filters at the light tight box and sample holder. The low frequency rejection is about $-120$ dB, even so the source power is dominated by long wavelength radiation for radiator temperatures below 11K, which can be seen in panel (c) of the figure. We decided not to increase the amount of filters as the power transmitted is below $10^{-22}$ $\mathrm{W}$, well below our detector detection threshold.

\section{Definition of Optical Efficiency in Multi-mode Detectors and its Estimation from Measured NEP}
\label{App_opt_coup}

In order to evaluate the optical coupling of a multi-mode detector system, the power absorbed by a detector in a measurement setup is characterised here. The power absorbed by the detectors when illuminated by the setup's blackbody radiator, modelled as a incoherent distributed source, is estimated as:

\begin{equation}
\label{Eq_P_abs_dis_comp}
P_{\mathrm{abs}} = \int_{f_1}^{f_2} \iint_{\Omega_s} B_s(f) A_{\mathrm{eff}}(f,\theta_{l}^{\mathrm{pw}},\phi_{l}^{\mathrm{pw}}) \mathrm{F_{il}}(f)\mathrm{d}f \mathrm{d}\Omega
\end{equation}

\noindent
where $B_s(f)$ is the brightness of the distributed source, $\mathrm{F_{il}}(f)$ is the response of the filter stages within the measurement setup; $f_1$ and $f_2$ are the lower and higher ends of the operation bandwidth, $\mathrm{d}\Omega=\sin{\theta_{l}^{\mathrm{pw}}}\mathrm{d}\theta_{l}^{pw} \mathrm{d}\phi_{l}^{pw}$and $\Omega_s$ is the solid angle of the source seen by the detector, see Fig. \ref{fig_SetupA}(a). $A_{\mathrm{eff}}(f,\theta_{l}^{\mathrm{pw}},\phi_{l}^{\mathrm{pw}})$ is the effective area of the detector:

\begin{equation}
\label{Eq_eff_area}
\begin{split}
 A_{\mathrm{eff}}& (f,\theta_{l}^{\mathrm{pw}},\phi_{l}^{\mathrm{pw}}) =  A_{\mathrm{lens}}  \big[\dfrac{1}{2}\eta_{ap}^{\mathrm{pol-1}}(f)F^{\mathrm{pol-1}}(f, \theta_{l}^{\mathrm{pw}},\phi_{l}^{\mathrm{pw}}) \\ +   &\dfrac{1}{2} \eta_{ap}^{\mathrm{pol-2}}(f)F^{\mathrm{pol-2}}(f, \theta_{l}^{\mathrm{pw}},\phi_{l}^{\mathrm{pw}})\big]
\end{split}
\end{equation}
where $A_{\mathrm{lens}}$ is the area of the lens aperture, $\eta_{ap}^{\mathrm{pol1-1}}$ and $\eta_{ap}^{\mathrm{pol1-2}}$  are the aperture efficiency of the detector for polarisation 1 and 2, respectively; and $F^{\mathrm{pol-1}}(f, \theta_{l}^{\mathrm{pw}},\phi_{l}^{\mathrm{pw}})$ and $F^{\mathrm{pol-2}}(f, \theta_{l}^{\mathrm{pw}},\phi_{l}^{\mathrm{pw}})$ are their normalised reception power patterns. 

\noindent

The spill over efficiency within  the measurement setup can be estimated as:
\begin{equation}
\label{Eq_SO_eff_setup}
 \eta_{\mathrm{so}}^{\Omega_s, \mathrm{pol-i}} (f) = \dfrac{\iint_{\Omega_s}F^{\mathrm{pol-i}}(f, \theta_{l}^{\mathrm{pw}},\phi_{l}^{\mathrm{pw}}) \mathrm{d}\Omega}{\iint_{2 \mathrm{\pi}}F^{\mathrm{pol-i}}(f, \theta_{l}^{\mathrm{pw}},\phi_{l}^{\mathrm{pw}}) \mathrm{d}\Omega }
\end{equation}

\noindent
 where the superscript ${\mathrm{pol-i}}$ represents the corresponding values for polarisation $i=1$ or $2$.
 
By using Eqs. (\ref{Eq_eta_f}) and (\ref{Eq_SO_eff_setup}), the power absorbed from a distributed source can be rewritten as:

\begin{equation}
\label{Eq_P_abs_dis_S2}
\begin{split}
&P_{\mathrm{abs}} = \int_{f_1}^{f_2} \dfrac{1}{2}\lambda_0^2 B_s(f) \mathrm{F_{il}}(f)\eta_{\mathrm{so}}^{\Omega_s,\mathrm{pol-1}} (f) \dfrac{\eta_{ap}^{\mathrm{pol-1}}(f)}{\eta_{f}^{\mathrm{pol-1}}(f)}  \mathrm{d}f \\ & +\int_{f_1}^{f_2} \dfrac{1}{2} \lambda_0^2 B_s(f) \mathrm{F_{il}}(f)\eta_{\mathrm{so}}^{\Omega_s,\mathrm{pol-2}} (f) \dfrac{\eta_{ap}^{\mathrm{pol-2}}(f)}{\eta_{f}^{\mathrm{pol-2}}(f)}\mathrm{d}f 
\end{split}
\end{equation}

\noindent
where the power absorbed from each polarisation is separated explicitly. The optical efficiency (also referred to as the normalised throughput in the literature \cite{Griffin2002}, \cite{Llombart2018}) for a multi-mode detector for each polarisation is then expressed as: 

\begin{equation}
\label{Eq_eta_op}
\eta_{opt}^{\mathrm{pol-i}}(f)=\dfrac{\eta_{\mathrm{so}}^{\Omega_s,\mathrm{pol-i}}(f)\eta_{ap}^{\mathrm{pol-i}}(f)}{\eta_{f}^{\mathrm{pol-i}}(f)}
\end{equation}

\noindent
Using this definition, Eq. (\ref{Eq_P_abs_dis_S2}) is rearranged  for each polarisation as:

\begin{equation}
\label{Eq_P_abs_dis_S3}
P_{abs}^{\mathrm{pol-i}}  =  \int_{f_1}^{f_2} \eta_{opt}^{\mathrm{pol-i}}(f) P_{sf}(f) \mathrm{d}f
\end{equation}
 \noindent
where  

\begin{equation}
\label{Eq_P_BB_one_pol}
P_{sf}(f)=\dfrac{1}{2}\lambda_0^2 B_s(f) \mathrm{F_{il}}(f)
\end{equation}

\noindent
is the blackbody power transmitted per frequency and per polarisation which can couple to a single mode detector.

The photon noise limited NEP of a KID is given by \cite{Baselmans2022}: 

\begin{equation}
\label{Eq_NEP_tot}
 \mathrm{NEP}_{\mathrm{tot}}^2 = \mathrm{NEP}_{\mathrm{Poisson}}^2 +\mathrm{NEP}_{\mathrm{Wave}}^2 +\mathrm{NEP}_{\mathrm{qp}}^2
\end{equation}

We need to evaluate these terms at the detector, i.e. with respect to absorbed power, as the bunching term is determined by the absorption process. The first righthand side term in Eq. (\ref{Eq_NEP_tot}), $\mathrm{NEP}_{\mathrm{Poisson}}$, is the Poisson noise due to random photon fluctuations which is the dominating term in the Wien's limit. This term is expressed using Eq. (\ref{Eq_P_abs_dis_S3}) and \cite{Ferrari2017} as:

\begin{equation}
\label{Eq_NEP_pos}
 \mathrm{NEP}_{\mathrm{Poisson}}^2(P_{\mathrm{abs}}^{\mathrm{pol-i}})  = \int_{f_1}^{f_2} 2\eta_{opt}^{\mathrm{pol-i}}(f) P_{sf}(f)hf \mathrm{d}f
 \end{equation}

 \noindent
where $f$ is the frequency of the arriving photons. The second righthand side term in Eq. (\ref{Eq_NEP_tot}), $\mathrm{NEP}_{\mathrm{Wave}}$, is the wave bunching term which dominates the noise in Rayleigh-Jeans’s limit. This term is expressed using Eq. (\ref{Eq_P_abs_dis_S2}) and \cite{Ferrari2017}  as:

\begin{equation}
\label{Eq_NEP_wave}
\begin{split}
   \mathrm{NEP}_{\mathrm{Wave}}^2& (P_{\mathrm{abs}}^{\mathrm{pol-i}}) =   \\  &\int_{f_1}^{f_2} 2(\eta_{opt}^{\mathrm{pol-i}}(f))^2 P_{sf} (f)hf \mathrm{F_{il}}(f) O_f(f)\mathrm{d}f
 \end{split}
 \end{equation}
 
 \noindent
where $O_f(f)=(e^{hf/(k_BT)}-1)^{-1}$ is the photon occupation number per mode, and $T$ is the temperature of the radiator. 

The final righthand side term in Eq. (\ref{Eq_NEP_tot}), $\mathrm{NEP}_{\mathrm{qp}}$, is the noise added due to random recombination of quasi-particles in a KID. This term is expressed using Eq. (\ref{Eq_P_abs_dis_S2}) and \cite{Ferrari2017} as:

\begin{equation}
\label{Eq_NEP_R}
 \mathrm{NEP}_{\mathrm{qp}}^2 (P_{\mathrm{abs}}^{\mathrm{pol-i}})  = \int_{f_1}^{f_2} 4\eta_{opt}^{\mathrm{pol-i}}(f) \Delta P_{sf}(f)/\eta_{\mathrm{pb}} \mathrm{d}f
 \end{equation}
 \noindent

 \noindent
where $\eta_{\mathrm{pb}}$ is the quasi-particle creation efficiency.

In the limit of small operational bandwidth around a central frequency $f_0$ defined by a set of Quasi-optical filters the absorbed power can be expressed by using Eq. (\ref{Eq_P_abs_dis_S3}) as:

\begin{equation}
\label{Eq_Pabs}
P_{abs}^{\mathrm{pol-i}}  =  \eta_{opt,f_0} \int_{f_1}^{f_2}P_{sf}(f) \mathrm{d}f 
\end{equation}
\noindent
 where $\eta_{opt,f_0}$ is the coupling efficiency at $f_0$ with respect to a single mode and a single polarisation.
 
The total photon noise limited NEP of the MKID as a function of the power absorbed can now be readily obtained by substituting Eqs. (\ref{Eq_NEP_pos}), (\ref{Eq_NEP_wave}) and (\ref{Eq_NEP_R}) in Eq. (\ref{Eq_NEP_tot}): 

\begin{equation}
\label{Eq_NEPph_abs}
\begin{aligned}
\mathrm{NEP}_{\mathrm{Ph}, P_{\mathrm{abs}}}^2 =&\\
& \eta_{opt,f_0} \int_{f_1}^{f_2}2P_{sf}(f) hf\mathrm{d}f + \\
& \eta_{opt,f_0}^2 \int_{f_1}^{f_2}2P_{sf}(f) hf\mathrm{F_{il}}(f) O_f(f) +\\
& \eta_{opt,f_0}\int_{f_1}^{f_2}4\Delta P_{sf}(f)/\eta_{pb}\mathrm{d}f 
\end{aligned}
\end{equation}

   \noindent
In an analogous way we can substitute Eq. (\ref{Eq_Pabs}) in Eq. (\ref{Eq_NEP_Ps}) to relate the experimental NEP as a function of source and absorbed power to each other:

\begin{equation}
\label{Eq_NEPph_s}
\mathrm{NEP}_{\mathrm{exp}, P_{\mathrm{abs}}}^2 = S_x (\dfrac{dx}{\mathrm{d}P_{\mathrm{abs}}})^{-2} = \eta_{opt}^2 \mathrm{NEP}_{\mathrm{exp}, P_{s}}^2 
\end{equation}

In our experiment we measure first the detector NEP as function of the source power available in front of the detector, Eq. (\ref{Eq_NEP_Ps}), which depends on the black body temperature and filters used in the experiment. To evaluate the NEP at the detector we need to obtain the optical efficiency $\eta_{opt}(f_0)$. This can be achieved in the limit that the detector NEP is photon noise limited. Following Janssen \emph{et al.} \cite{Janssen2013} this is the case when the KID noise power spectral density is white with a roll-off given by the quasiparticle lifetime. In our experiment this is the case for black body temperatures equal to and exceeding $18$ K, as can be seen from Fig. \ref{fig_ResultA2}. In these cases  $\mathrm{NEP}_{\mathrm{Ph}, P_{\mathrm{abs}}}=\mathrm{NEP}_{\mathrm{exp},P_{abs}}$ and we can use Eqs. (\ref{Eq_NEPph_abs}) and (\ref{Eq_NEPph_s}) to obtain the optical efficiency with respect to a single mode and a single polarisation: 
 

\begin{equation}
\label{Eq_eta_m}
\eta_{opt,f_0} =
\frac{ \int_{f_1}^{f_2}2P_{sf}(f) hf\mathrm{d}f + \int_{f_1}^{f_2}4\Delta P_{sf}(f)/\eta_{pb}\mathrm{d}f }   {\mathrm{NEP}_{\mathrm{exp},P_s}^2 - \int_{f_1}^{f_2}2P_{sf}(f) hf\mathrm{F_{il}}(f) O_f(f) }
\end{equation}

   \noindent
which can be evaluated experimentally as all its terms are known. 

For a dual polarised detector Eq. (\ref{Eq_Pabs}) is valid when:

\begin{equation}
\label{Eq_eta_opt_tot}
\eta_{opt, f_0} = \eta_{opt} ^{\mathrm{pol-1}}(f_0) + \eta_{opt} ^{\mathrm{pol-2}}(f_0)
\end{equation}

In this work, the results shown for the optical efficiency compare the measured response using Eq. (\ref{Eq_eta_m}) against the ones estimated by the model using Eqs. (\ref{Eq_eta_opt_tot}) and (\ref{Eq_eta_op}). 
 
 
\section{Additional Experimental Results}
\label{App_Result}

\begin{figure*}[!t]
 \includegraphics[width=1\textwidth]{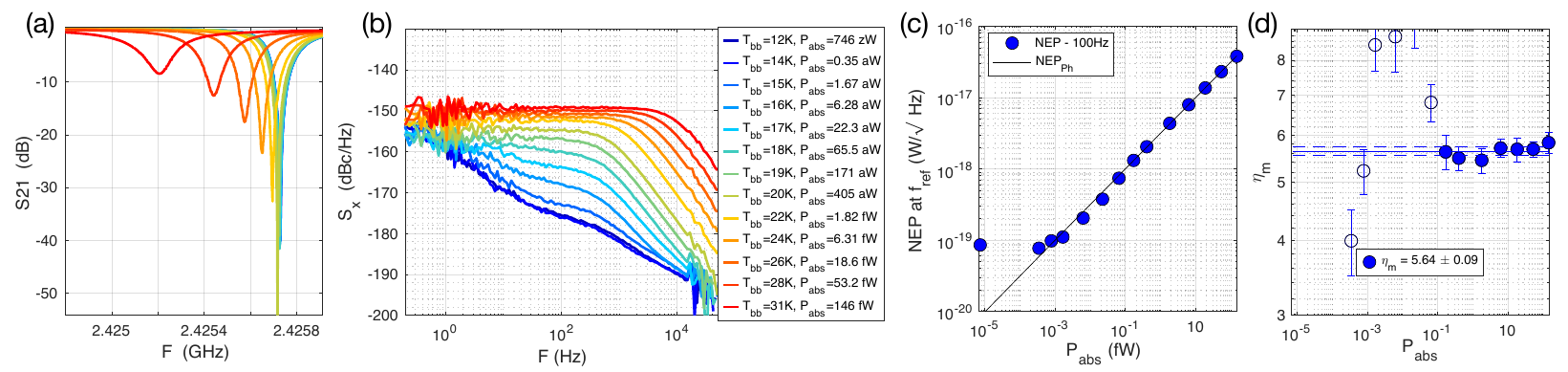}
\caption{Results of an experiment with $12$ radiator temperatures for KID $9$ with $F_\mathrm{{res}} =2.43$ GHz. (a) Forward transmission for all measured radiator temperatures, clearly showing the KID dip reducing and moving to lower frequencies. (b) Reduced frequency noise spectra. (c) NEP spectra, where we clearly observe the transition from a (partial) detector/1f noise limited NEP to a white, photon noise limited NEP for $P_{\mathrm{abs}}\mathrm{>68\;fW}$. (d) NEP as function of absorbed power at a modulation frequency of $F_{\mathrm{mod}}=100$ Hz.  }
\label{fig_ResultA2} 
\end{figure*}

To get more insight in the radiation power dependence of the KID noise and NEP, we measured for two KIDs the NEP for $12$ radiator temperatures, the (representative) result for KID $9$ is shown in Fig. \ref{fig_ResultA2}. In panel (a) we show the KID resonance features, where we clearly see the expected frequency shift and depth decrease with increasing absorbed power. Interestingly at a radiator temperature of $31$ K, corresponding to $P_{\mathrm{abs}}\mathrm{=151\;fW}$ we still observe an $8$ dB deep resonance, indicating that this KID works well at power levels exceeding $151$ $\mathrm{fW}$. In panels (b) and (c) we show the noise and NEP respectively. It is clear that the device moves from partly 1/f noise limited to background limited performance, with a white noise down to $0.1$ Hz, for radiator temperatures exceeding $18$K, corresponding to $P_{\mathrm{abs}}\mathrm{=65\;aW}$. 

In Fig. \ref{fig_ResultA2} (c) we give the NEP at $100$ Hz modulation frequency, which shows  clearly how the NEP follows the expected square root power dependence for background limited operation down to $P_{\mathrm{abs}}\approx\mathrm{1\;aW}$, indicating the superior performance at higher modulation frequencies. The inset shows the measured optical efficiency for each radiation power. To calculate the efficiency we only use the data with power levels exceeding $0.1$ $\mathrm{fW}$ where, at $100$ Hz, the device is background limited. The KIDs in this chip all show a very similar NEP under dark conditions and at $130$ mK, this is shown in Fig. \ref{fig_ResultA1}.

\begin{figure}[!t] 
\centering
\includegraphics[width=3in]{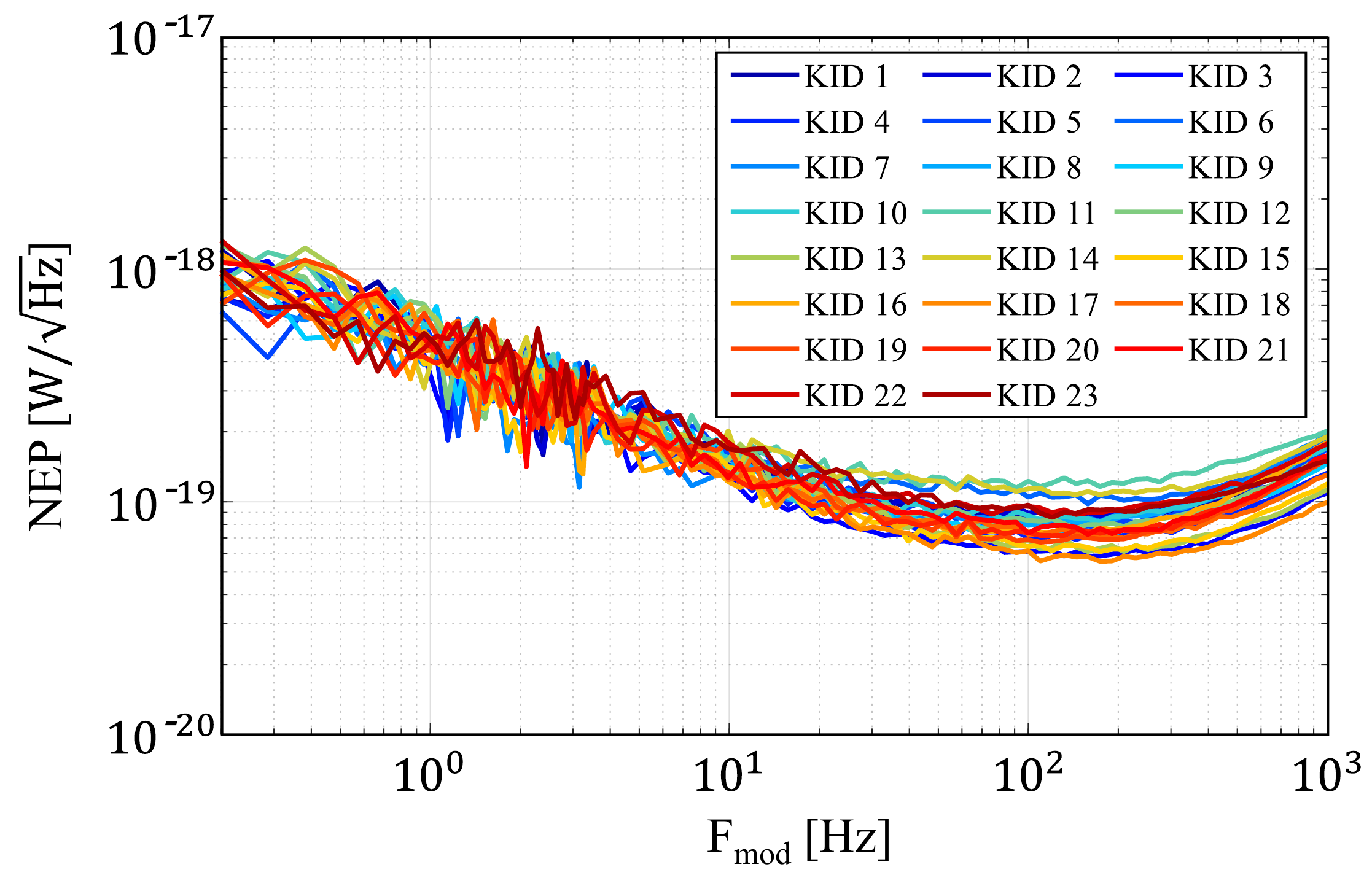}
\caption{NEP spectra for all KIDs at negligible radiation load and at $130$ mK.}
\label{fig_ResultA1} 
\end{figure}

In Fig. \ref{fig_Result_opt_4.5mm} the optical efficiencies are shown for all $23$ KIDs which are obtained experimentally when the diameter of the aperture between the black body radiator and the lens array is reduced from $15$ to $4.5$ $\mathrm{mm}$ with respect to the data provided in Fig. \ref{fig_Result_eta_opt_15} (corresponding to a full width opening angle of $4.8$$^\circ$). The nominal and misaligned theoretical values of the optical efficiency are also calculated and compared against experimental ones showing a very good agreement.

\begin{figure}[!t]
\centering
\includegraphics[width=3in]{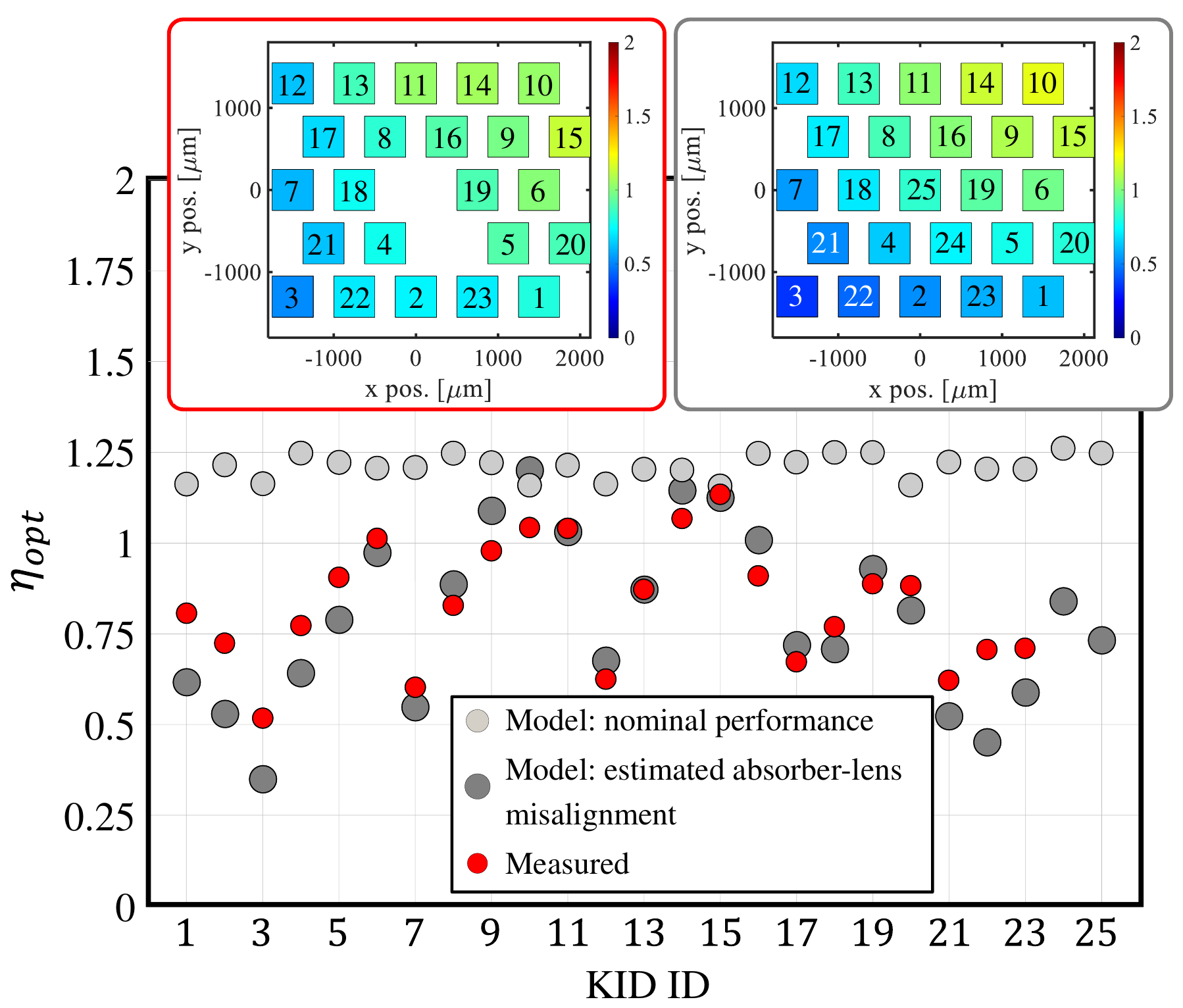}
\caption{Optical efficiency for all KIDs when the diameter of the aperture is reduced from $15$ $\mathrm{mm}$ (case in Fig. \ref{fig_Result_eta_opt_15}) to $4.5$ $\mathrm{mm}$. The left and right-hand side insets show the spatial dependency of the optical efficiency for each KID from the measured and modelled data, respectively, where the KID's IDs are also provided. }
\label{fig_Result_opt_4.5mm} 
\end{figure}

}

\bibliographystyle{IEEEtran}

\bibliography{lens_abs_v5.2}

 
\end{document}


\section*{Supplementary: Tolerance Study in Lens Absorbers}
\label{App_tol}

In this Appendix, the degradation of the lens absorber performance in terms of aperture efficiency due to fabrication inaccuracies are investigated and compared to the nominal performance. The nominal lens absorber designs are the ones discussed in Sec. III of the accompanying manuscript operating at two central frequencies of $6.98$ and $12$ THz.

\subsection*{Deviations in the lens radii of curvature }

A nominal elliptical lens element has an eccentricity, $e$, related to the contrast between inside and outside of its boundaries, namely here: $e = 1/\sqrt{\varepsilon_r}$ when $\varepsilon_r$ is the permittivity of the silicon dielectric. Limitations in fabrication lead to deviations of the curvature and imperfect focusing capabilities. Curvature errors in the fabrication of an integrated lens are typically reported in terms of deviations from the nominal radii of curvature of the surface. Such a deviation is reported as $\Delta R/R^{\mathrm{nom}}$, where $R^{\mathrm{nom}}$ is the nominal radius of the curvature and $\Delta R = R^{\mathrm{er}} - R^{\mathrm{norm}}$ is the deviation from the nominal value, and $R^{\mathrm{er}}$ is the imperfect radius. Here, following realistic fabrication scenarios, such an error is modelled as a lens element with the nominal aperture diameter and apex to focus length while the curvature is modified. 

\begin{figure}[!t]
\centering
\includegraphics[width=3in]{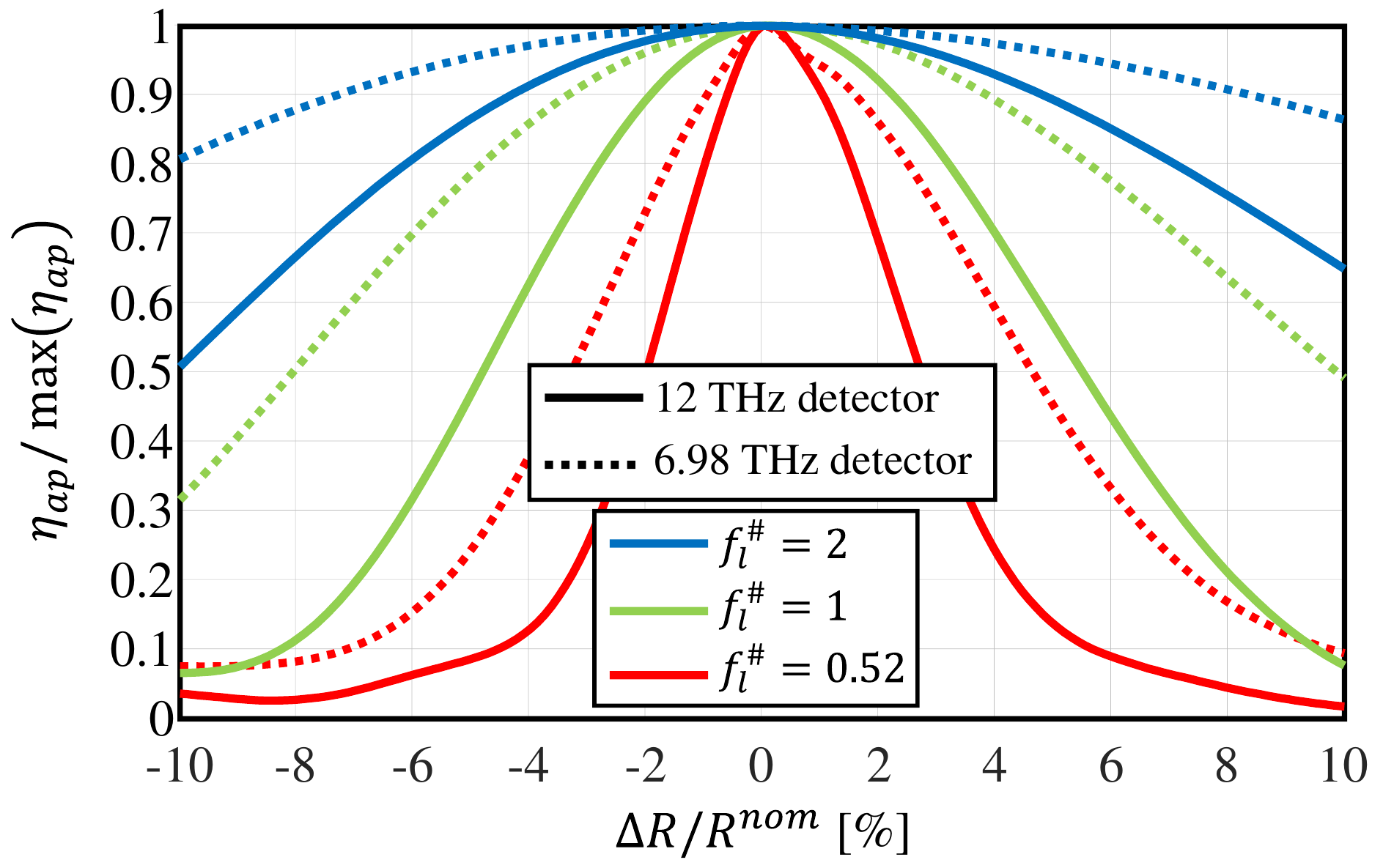}
\caption{Relative degradation of the aperture efficiency for $6.98$ and $12$ THz absorbers below three lens f-numbers. The absorber's unit cell designs are reported in Sec.III while the sampling rate for each lens cases is maintained as $w_{x/y}\simeq 2.55 \lambda_d f_{l}^{\#}$. }
\label{fig_curve_radii_er} 
\end{figure}

The degradation of aperture efficiency of the proposed absorber designs in Sec. III below three lens f-number cases is shown Fig. \ref{fig_curve_radii_er}. The absorber side lengths $w_x$ and $w_y$ are scaled for each lens component to have the same absorber sampling rate, $w_{nx/y}=w_{x/y}/(\lambda_d f_l^{\#})$, between all cases. As it can be seen, despite the fact that the absolute radii error is larger when the lens f-number is larger, such lens absorbers are significantly more robust against lens surface errors. This behaviour is due to the smaller subtended angle of larger lens f-numbers which leads to lesser distortion of absorber's response. As expected, the absorbers operating at higher frequencies are more sensitive to the same curvature error. It is also worth noting that due to the incoherent nature of the absorbers, sensitivity of their performance is significantly lower to curvature errors with respect to a lens antenna detector as described in \cite{Pascual2025}.

\subsection*{Surface roughness of the lens}

Ruze attributes the reduction of the gain in reflector antennas in transmission mode to random phase errors caused by the reflector surface roughness \cite{Ruze1966}. Such a representation is modelled here in reception to modify the direct PWS arriving to absorber by adding a random phase error:

\begin{equation}
\label{Ruze1966_rx}
\vec{E}_{d,\mathrm{er}}^{\mathrm{PWS}} = \vec{E}_{d,\mathrm{nom}}^{\mathrm{PWS}} e^{-\mathrm{j} k_{\mathrm{avg}}d_{\mathrm{rough}}}
\end{equation}

\noindent
where $\vec{E}_{d,\mathrm{nom}}^{\mathrm{PWS}}$ and $\vec{E}_{d,\mathrm{er}}^{\mathrm{PWS}}$ are the nominal and with surface roughness direct PWSs, respectively; $k_{\mathrm{avg}}= (k_0+k_d)/2$ is the averaged propagation constant between the one of free space, $k_0$ and silicon dielectric $k_d$; $d_{\mathrm{rough}}$ is a random surface modification with a root mean square (RMS) value of $\epsilon_{\mathrm{rms}}$. 
\begin{figure}[!t]
\centering
\includegraphics[width=3in]{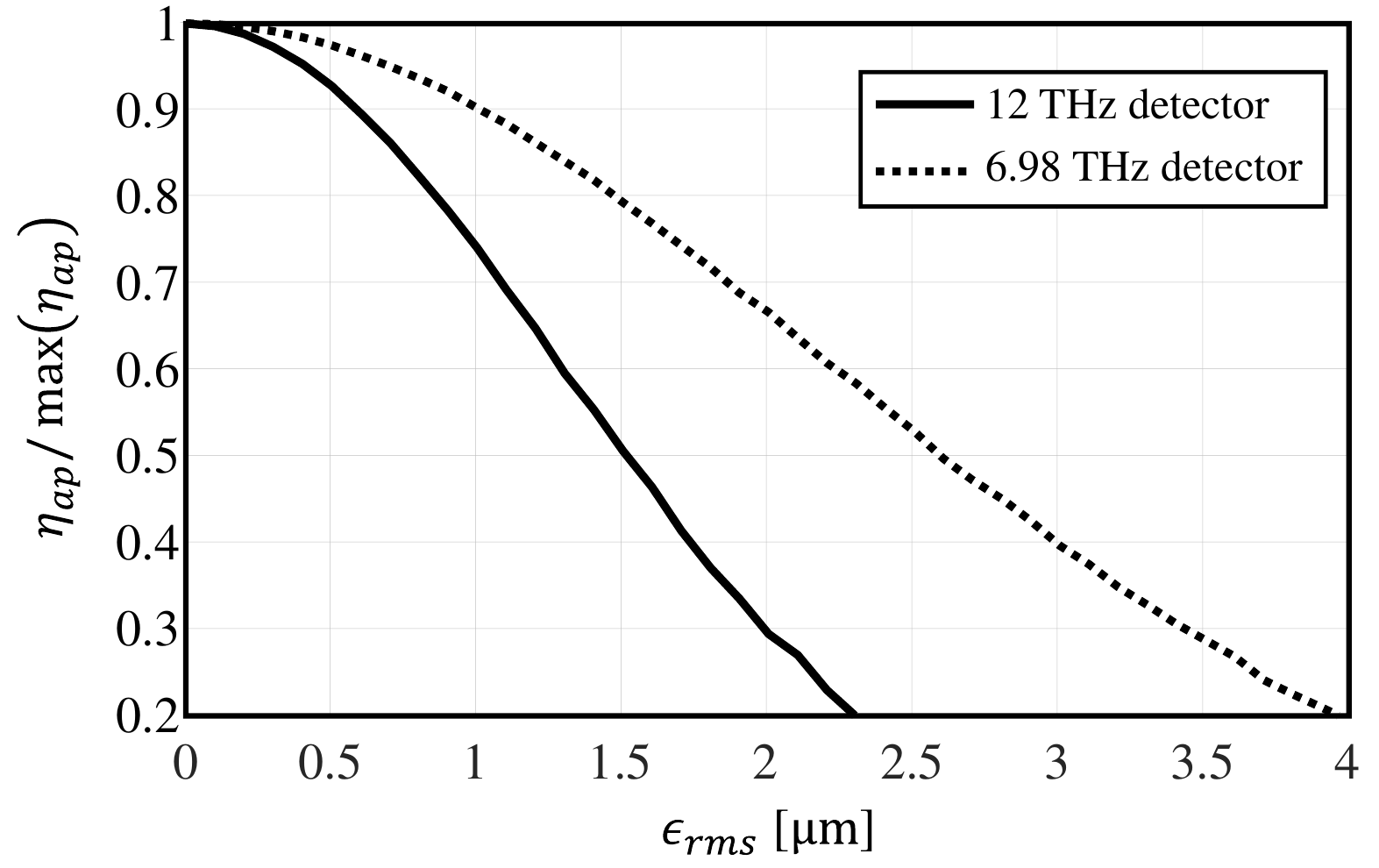}
\caption{Relative degradation of the aperture efficiency for 6.98 and 12 THz absorbers below lens elements with a surface roughness indicated by the RMS value of $\epsilon_{\mathrm{rms}}$.}
\label{fig_surf_rough_er} 
\end{figure}

The degradation of aperture efficiency with respect to the RMS surface error is reported in Fig. \ref{fig_surf_rough_er}. The results obtained are independent of the lens f-number. The loss in performance shown here is in-line with the equivalent study for lens antennas in \cite{Pascual2025}. In the referenced work, this loss is represented as Ruze's formula for transmitting surfaces: $e^{- \frac{2\mathrm{\pi}\epsilon_{\mathrm{rms}}}{\lambda_{\mathrm{avg}}} }$ where $\lambda_{\mathrm{avg}} = (\lambda_0+\lambda_d)/2$ is the averaged wavelength between the two media at the boundaries of the lens surface. As a result, the sensitivity of an incoherent detector to surface roughness is not lower with respect to an antenna system.    


\subsection*{Alignment of the lens and the absorber}

At FIR wavelengths, a few microns of lateral misalignment during the assembly phase between the detector and lens focus is a significant value in terms of wavelength. This situation leads to extreme alignment requirements for lens antenna systems above $7$ THz \cite{Pascual2025}. Moreover, height variations within a silicon wafer leads to significant antenna phase errors. This height imperfections are referred here as a vertical misalignment.

\begin{figure}[!t]
\centering
\includegraphics[width=3in]{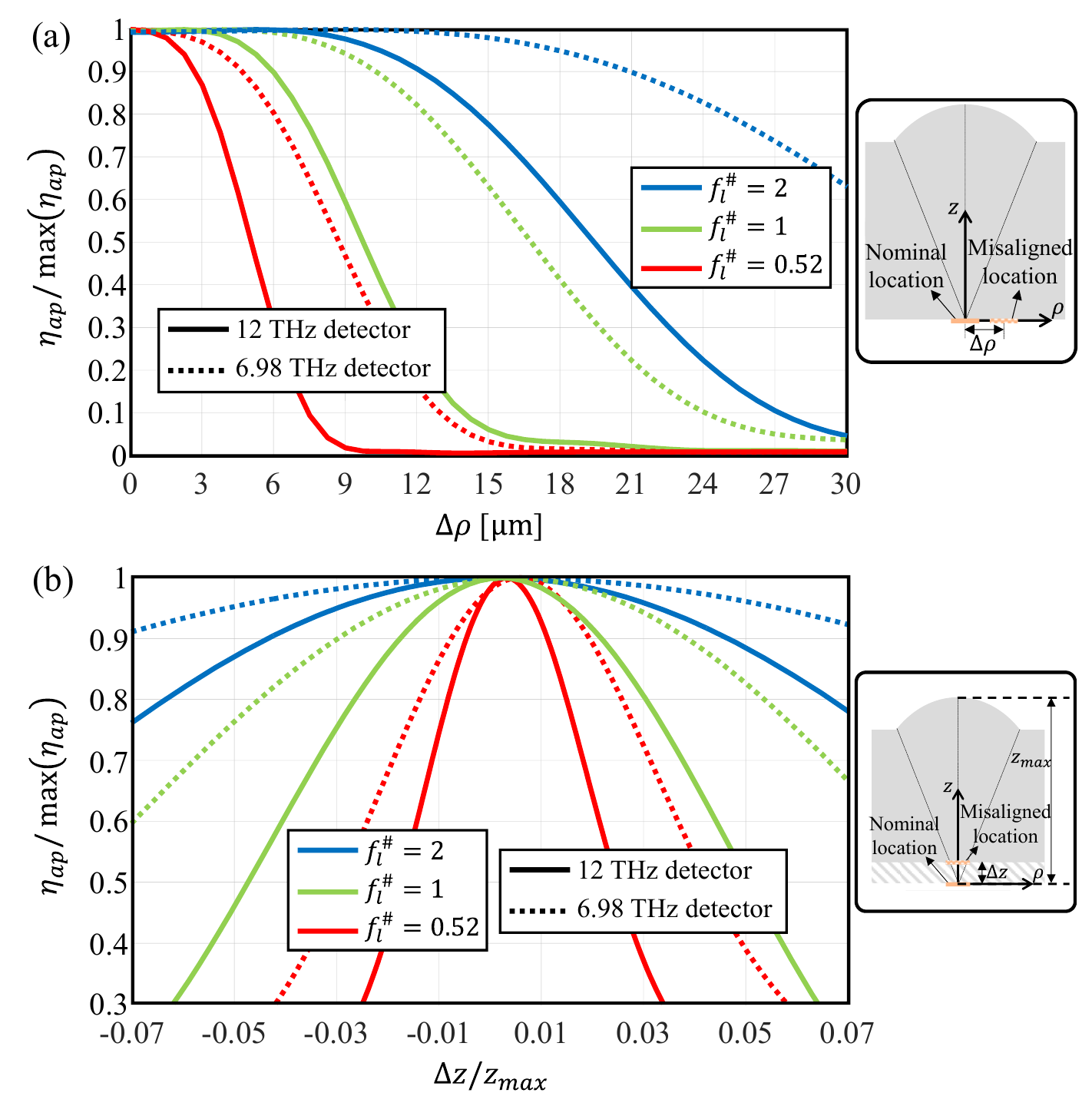}
\caption{Relative degradation of the aperture efficiency for $6.98$ and $12$ THz absorbers below lens elements with (a) lateral and (b) vertical misalignments. The inset illustrates the scenarios considered. The vertical misalignment is normalised to the focal distance of the lenses, $z_{\mathrm{max}}$, which is equal to $1419$, $739$, and $475$$\mathrm{\mu m}$ for lenses with diameter of $700\mathrm{\mu m}$ and $f_l^{\#}$ of $2$, $1$, and $0.52$, respectively.}
\label{fig_mis_allign_er} 
\end{figure}

 Fig. \ref{fig_mis_allign_er} indicates a minute sensitivity to alignment issues for lens absorbers in comparison to an antenna system. Similar to surface curvature error case, absorber is significantly more robust due to its incoherent nature of detection. 
       In a first order view, focal fields of the three lens elements are equivalent to one another but scaled proportional to their f-numbers. As a result, a same absolute lateral misalignment is equivalent to missing a larger portion of the focal field in a lens with a smaller f-number. Moreover, a lens with larger f-number has a more extended depth of focus leading to high absorption efficiency even when the absorber is slightly above or below the nominal focus point.  

\subsection*{Presence of an air gap in the assembled device}

\begin{figure}[!t]
\centering
\includegraphics[width=3in]{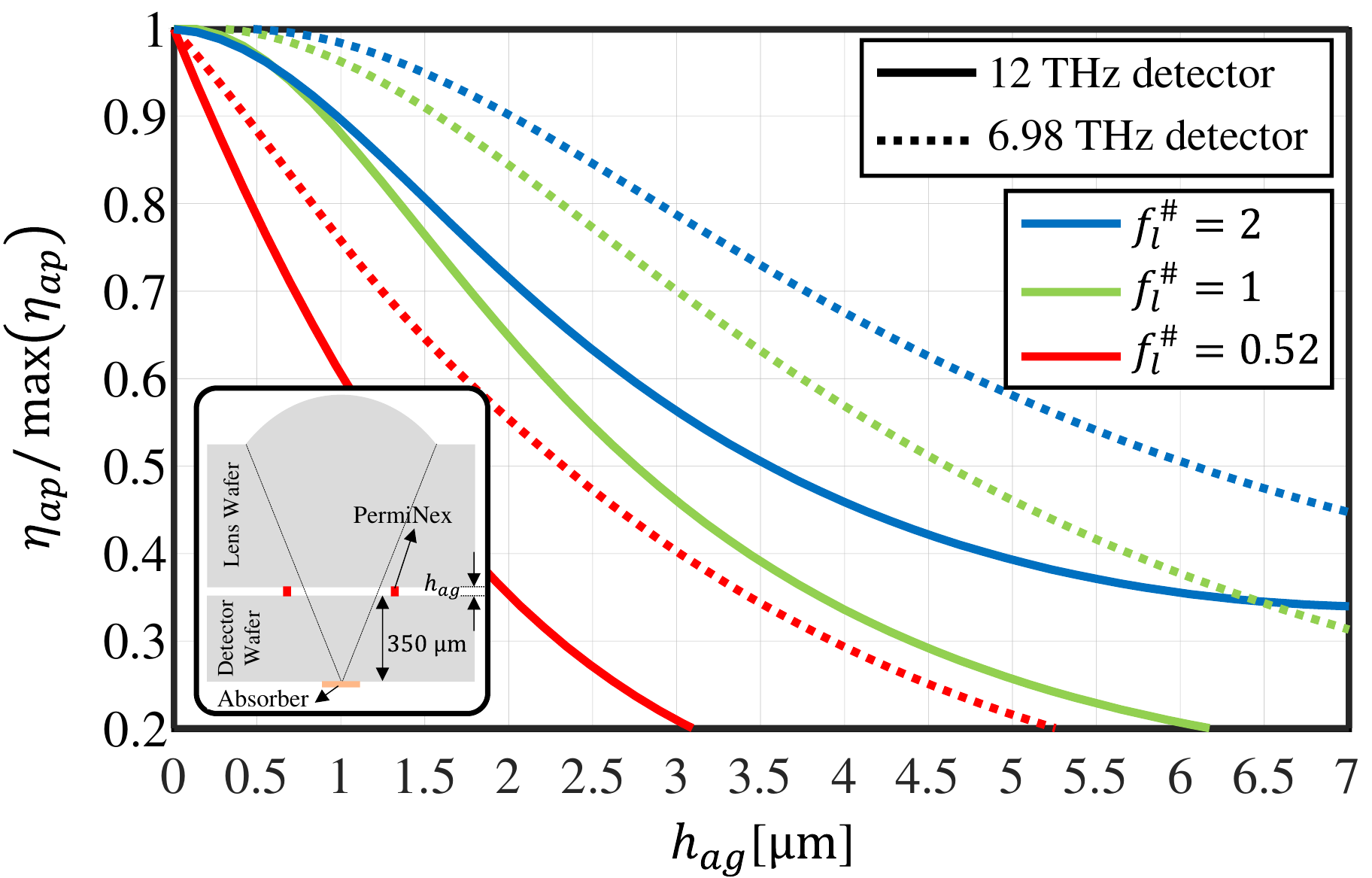}
\caption{Relative degradation of the aperture efficiency for $6.98$ and $12$ THz absorbers due to the presence of an air gap between silicon wafers. The inset illustrates the scenarios considered.  }
\label{fig_air_gap} 
\end{figure}

In this work, our lens and KID devices are fabricated on separate wafers and assembled using Perminex, which is present only at locations without any EM field, hence for the EM coupling an air gap equal to the Perminex thickness is created. The two wafers are shown in the inset of Fig. \ref{fig_air_gap}. Fig. \ref{fig_air_gap} reports the loss in aperture efficiency as a function of the air gap thickness when it is located at $350$ $\mathrm{\mu m}$ above the absorber plane, as is the case in this work. The performance is evaluated using the described Floquet wave circuit where the air gap is modelled as an extra transmission line with the proper thickness at the expected location. As can be seen, controlling the thickness of this air gap is critical in order to preserve the lens absorbers performance. Moreover, since for a skewed incident plane wave, a vertical discontinuity within the propagation medium is effectively extended, the absorber's plane wave response to larger incident angles is more disturbed by the presence of an air gap. As a result, for a lens with a larger angular region (lower lens f-number), the aperture efficiency is more sensitive to the air gap thickness, as can be clearly seen in the figure.

\subsection*{Estimated tolerances in the fabricated devices}

In the case of the fabricated lens array, operating at $6.98$ THz, a 3D surface analysis led to the following estimates. For the lens surface quality, a curvature error of $\Delta R/R^{\mathrm{nom}}=-0.7\%\pm2.4\%$, with a RMS surface error of $\epsilon_{\mathrm{rms}}<1$ $\mathrm{\mu m}$ is estimated. Moreover, a negligible vertical misalignment between absorber plane and lens focus is expected. The effect of these small lens related inaccuracies on the device performance is neglected in the modelled optical efficiency. On the other hand, as discussed in Sec. IV-B, a lateral misalignment of  $\simeq 29.9\hat{x} + 15.1\hat{y}$ $\mathrm{\mu m}$ between the lens and absorber centres is expected and included in the modelling of the performance. This misalignment led to a tilt in the reception pattern of the lens absorbers as shown in the inner inset of Fig. 13 of the accompanying manuscript. Finally, based on the Perminex bonding procedure, an air gap of $h_{ag}\simeq 0.5$ $\mathrm{\mu m}$ is enforced between the detector and the lens wafers leading also to negligible performance penalties. 

\bibliographystyle{IEEEtran}

\bibliography{Supplementary_tolerance_study}